\documentclass[12pt]{iopart}
\usepackage[caption=false, subrefformat=parens,labelformat=parens]{subfig}
\usepackage[backend=bibtex, style=phys, articletitle=true,biblabel=brackets,%
chaptertitle=true,pageranges=true]{biblatex}
\DeclareFieldFormat[article]{title}{\mkbibemph{#1}}
\DeclareFieldFormat[inproceedings]{title}{\mkbibemph{#1}}
\DeclareFieldFormat[Inbook]{title}{\mkbibemph{#1}}
\usepackage{graphicx}
\usepackage{bbold}
\usepackage{url}
\usepackage{booktabs}
\usepackage{iopams}
\usepackage[percent]{overpic}
\usepackage[normalem]{ulem} 
\usepackage[dvipsnames]{xcolor}

\usepackage{appendix}

\renewcommand{\theequation}{I.\arabic{equation}}

\usepackage{tikz}
\usetikzlibrary{decorations.shapes}
\tikzset{decorate sep/.style 2 args=
{decorate,decoration={shape backgrounds,shape=circle,shape size=#1,shape sep=#2}}}

\usepackage{hyperref}

\begin{document}

\title[Multivariate Distributions in Non--Stationary Complex Systems I]
{Multivariate Distributions in Non--Stationary Complex Systems I:
Random Matrix Model and Formulae for Data Analysis}

\author{Efstratios Manolakis\footnote{Now at: Dipartimento di Fisica e Astronomia Ettore Majorana, Universit\`a degli Studi di Catania, and
Dipartimento di Fisica e Chimica Emilio Segr\`e, Universit\`a degli Studi di Palermo, Italy}, Anton J.~Heckens, Benjamin Köhler and Thomas Guhr}
\address{Fakult\"at f\"ur Physik, Universit\"at Duisburg--Essen, Duisburg, Germany}
\ead{efstratios.manolakis@phd.unict.it, anton.heckens@uni-due.de, benjamin.koehler@uni-due.de and thomas.guhr@uni-due.de}

\begin{abstract}
Risk assessment for rare events is essential for understanding
systemic stability in complex systems. As rare events are
typically highly correlated, it is important to study
heavy--tailed multivariate distributions of the relevant
variables,
especially in the presence of non--stationarity.
We use a generalized scalar product between correlation matrices to clearly demonstrate this non--stationarity. Further, we present a model that we recently put forward, which captures how the non--stationary fluctuations of correlations make the tails of multivariate distributions heavier. Here, we provide the resulting formulae including Gaussian or Algebraic features. Compared to our previous results, we manage to remove in the Algebraic cases one out of the two, respectively three, fit parameters which considerably facilitates applications. We demonstrate the usefulness of these results by deriving joint distributions for linear combinations of amplitudes and validating them with financial data. Furthermore, we explicitly work out the moments of our model distributions. In a forthcoming paper we apply the model to financial markets.
\end{abstract}

\section{\label{sec:level1}Introduction}

Ever more high--quality data accumulated in complex systems of all
kinds become available and trigger the need for a better understanding
and a quantitative modeling
\cite{mantegna1999introduction,Kutner_2019}. The data are typically
highly correlated, implying that a univariate data analysis is
insufficient.  Rare events in the often heavy tails of the
distributions are especially sensitive for the systemic risk and the
stability of a system. Another important aspect of complex systems is
their non--stationarity
\cite{Plerou_2002,munnixIdentifyingStatesFinancial2012,Wang_2020,wang2021collective,Bette_2023}.
Finance is a good example, but certainly not the only one. The
standard deviations or volatilities which are important statistical
estimators fluctuate seemingly erratically over time
\cite{schwert1989,Mandelbrot1997,Bekaert2000,BEKAERT2014,MAZUR2021}.
The mutual dependencies such as Pearson correlations or copulas
\cite{Pearson_1900,Sklar1959:Fonctions,sklar1973random,nelsen2006introduction,MUNNIX2011,salinas2019high}
which measure the relations within the financial markets show
non--stationarity variations as well which plays a particularly important
role in states of
crisis~\cite{munnixIdentifyingStatesFinancial2012,Chetalova_2015,Chetalova_2015_2,Stepanov_2015_MultiAsset,Stepanov_2015,Rinn_2015,Pharasi_2018,Heckens_2020,heckens2021new,heckens2022new,marti2021review,pharasi2020market,pharasi2021dynamics,James_2022,Wand_2023,Hessler_2023,Wand_2023_2}.

Our goal is, for complex systems in general, to assess and quantify
non--stationarity and to provide analytical model descriptions for the
multivariate distributions.  In
Refs.~\cite{Schmitt_2013,Schmitt_2014,schmitt2015credit,chetalova2015portfolio,Meudt_2015,Sicking_2018,Muehlbacher_2018}
we developed a model for the multivariate distributions in the context
of credit risk and portfolio optimization. We recently considerably
extended it \cite{Guhr_2021} to also properly capture algebraic
tails. Here, we present these results in a form directly applicable to
data.  The model is based on a separation of time scales, guided by
the observation that the effects due to non--stationarity accumulate
as the length of the considered time intervals increases. We assume a
certain behavior, for example approximate stationarity, within short
epochs and fluctuating correlations from epoch to epoch. Modeling the
latter with random
matrices~\cite{Mehta1967,Guhr_1998,GuptaNagar2000,Potters2021}, we are
able to provide analytical formulae with few parameters for the
multivariate distributions in the presence of
non--stationarity. Importantly, compared to our formulae in
Ref.~\cite{Guhr_2021}, we manage to reduce the number of fit
parameters in the algebraic cases from two to one or three to two,
respectively, which is highly useful for applications. As an example,
we show how to apply these results to combinations of amplitudes.  We
also provide new results on moments. From a formal mathematical point
of view, our random--matrix model is a matrix--valued extension of
compounding \cite{Dubey1970} or mixture \cite{BarndorffNielsen_1982}
approaches in statistics, but in contrast to these results, we are on
phenomenologically solid grounds. We model a truly existing ensemble
of empirical correlation matrices by an ensemble of random matrices.
Among many other things, our random--matrix model also gives a
justification and interpretation of single--variate ad--hoc approaches
\cite{Dubey1970,BarndorffNielsen_1982,BECK2003,Abul-Magd_2009,Doulgeris2009,Forbes2014}.
In a forthcoming paper~\cite{Heckens_PaperII} henceforth referred to
as II, we will present a careful comparison of financial data analysis
with the analytical model. In course of doing so, we will explain and
demonstrate in detail how to determine the parameters of the model
distributions.

The paper is organized as follows. In Sec.~\ref{sec:RMModel}, we give
an overview of our random matrix approach and bring the resulting
multivariate distributions in forms directly applicable to data. We
also calculate the moments of these four model distributions.  We give
our conclusions in Sec.~\ref{sec:Conclusion}.

\section{\label{sec:RMModel}Random Matrix Model for Multivariate Distributions}

In Sec.~\ref{sec:IdeaAndConcept}, we present the salient features of
the random matrix model. The process of rotation and aggregation for
the analytical distributions with arbitrary kinds of amplitudes is
explained in Sec.~\ref{subsec:agg}. In
Secs.~\ref{sec:RMTAmplitudeDist} and \ref{sec:RMTEnsembleDist}, we
specify two forms of multivariate distributions for the epochs and
calculate those on the long interval by employing two forms of random
matrix ensembles to model the non--stationarity. We arrive at four
ensemble averaged multivariate amplitude distributions, described in
Sec.~\ref{sec:RMTEnsembleAverageDist}. In
Sec.~\ref{sec:GeneralLinearComb}, we demonstrate how our results can
be used to obtain distributions for arbitrary combinations of
amplitudes, explicitly we focus on linear combinations.  
We calculate
the moments of the distributions in Sec.~\ref{sec:MomentsModelDist}.

\subsection{\label{sec:IdeaAndConcept}Idea and Concept}

\begin{figure}[htbp]
\captionsetup[subfigure]{labelformat=empty}
\centering
\begin{minipage}{0.7\textwidth}
\subfloat[]{\begin{overpic}[width=1.\linewidth]{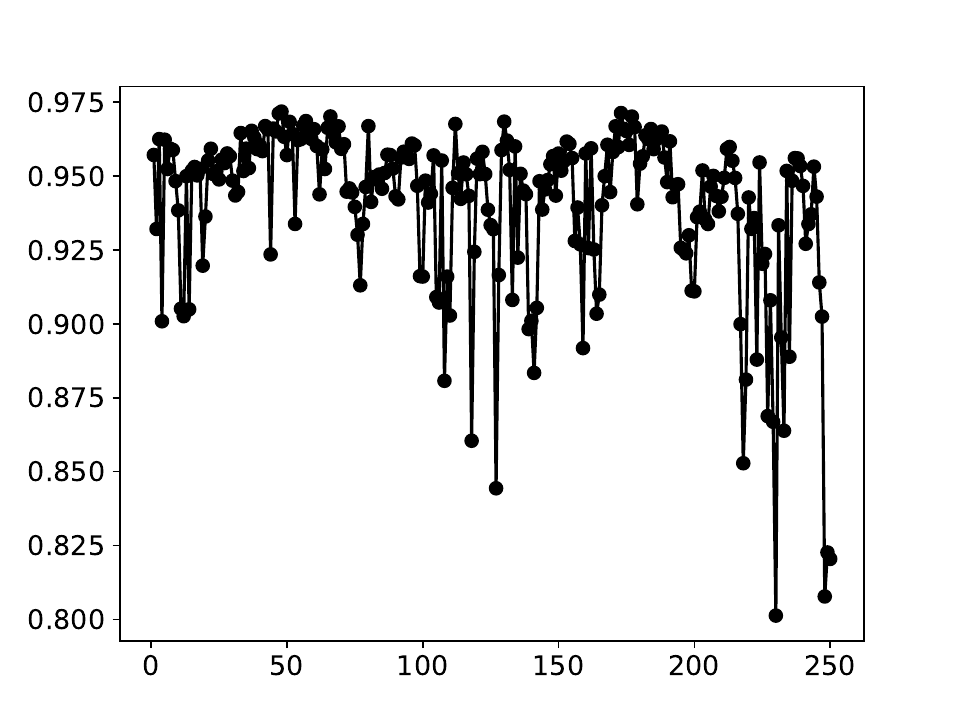}
\put(50,0){\makebox(0,0){\small\sffamily \large number of epochs $\mathsf{i}$}}
\put(0.2,35){\makebox(0,0){\rotatebox{90}{\small\sffamily \large $\cos \widetilde{\alpha_{i}}$}}}
\end{overpic}
}
\end{minipage}%
\caption{\label{fig:twocorr_4}Similarity measures $\cos \widetilde{\alpha}_{i}$ over the number of epochs $i$.}
\end{figure}

\begin{figure}[htbp]
\captionsetup[subfigure]{labelformat=empty}
\centering
\begin{minipage}{0.7\textwidth}
\subfloat[]{
\includegraphics[width=1.0\textwidth]{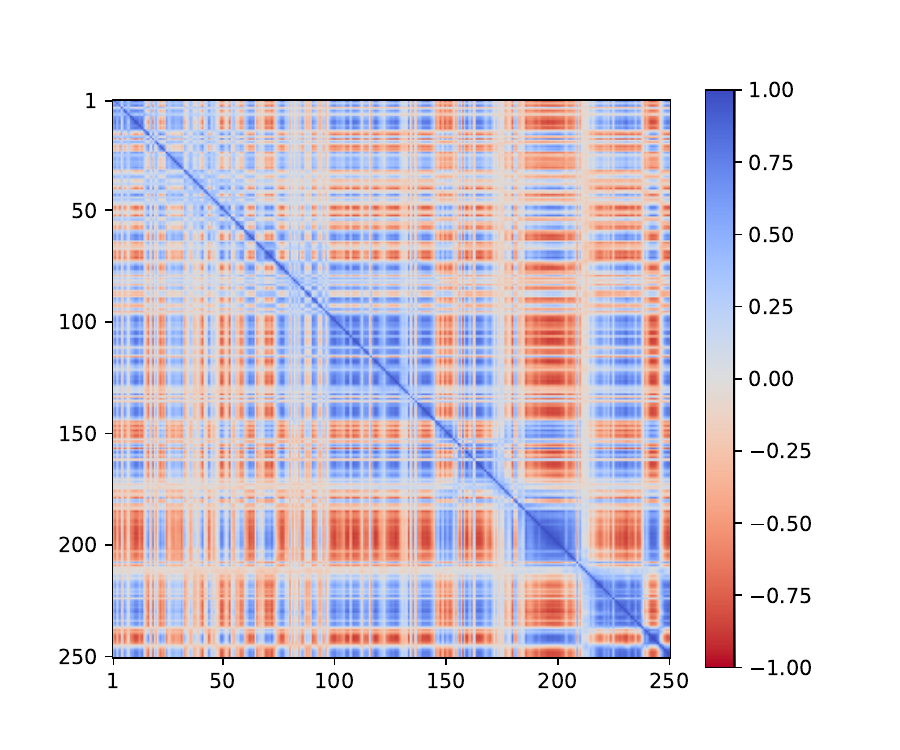}
}
\end{minipage}
\caption{\label{fig:twocorr}Non--stationarity in
the example of finance. Matrix of the similarity measure
$\cos \alpha_{ij}$ for the $N_{\mathrm{ep}}=250$ epochs of
the year 2014.}
\end{figure}

\begin{figure}[htbp]
\captionsetup[subfigure]{labelformat=empty}
\centering
\begin{minipage}{0.7\textwidth}
\subfloat[]{\begin{overpic}[width=1.\linewidth]{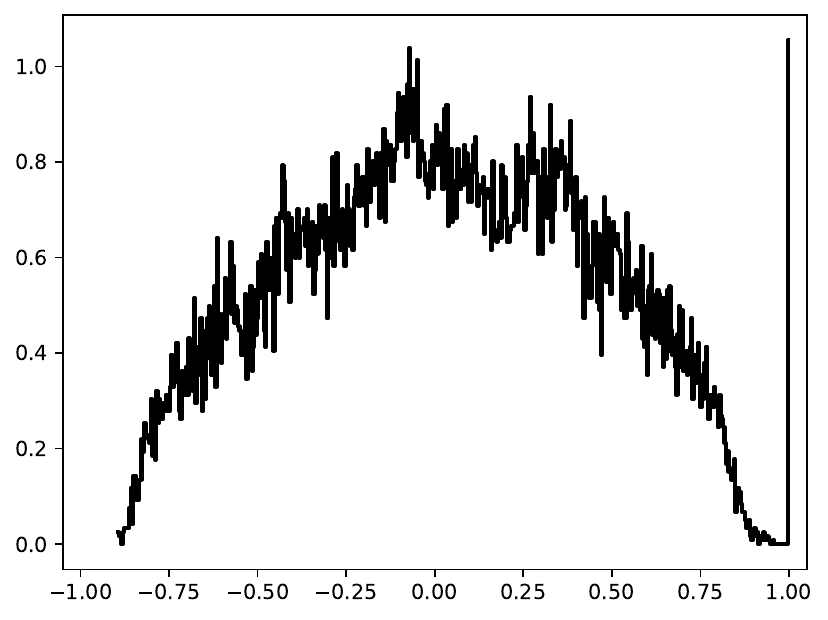}
\put(50,-2){\makebox(0,0){\small\sffamily \large  $\cos \alpha_{ij}$}}
\put(-2,35){\makebox(0,0){\rotatebox{90}{\small\sffamily \large pdf}}}
\end{overpic}
}
\end{minipage}
\caption{\label{fig:twocorr_2}Distribution of the similarity measure $\cos \alpha_{ij}$ for the year 2014.}
\end{figure}

Non--stationarity is ubiquitous in complex systems. Finance provides
good examples.  Correlation coefficients between different stocks vary
when analyzed in a sliding sample window.  There is no reason for them
to be constant, as the business relations, the company performances,
the traders' market expectations and so on change in time. This
prompts us to treat non--stationarity of complex systems in general in
the following way. To model multivariate distributions of $K$
amplitudes $r_k, \ k=1,\ldots,K$ ordered in a vector
$r=(r_1,\ldots,r_K)$ on a long time interval, we account for
fluctuating correlations by separating the time scales as in
Fig.~\ref{fig:scalesep} into $N_{\mathrm{ep}}$, say, epochs. In each
\begin{figure}[hbtp]
\centering
\begin{tikzpicture}
\draw[ultra thick] (0,0.5) -- (0,-0.5);
\draw[ultra thick] (2,0.5) -- (2,-0.5);
\draw[ultra thick] (4,0.5) -- (4,-0.5);
\draw[ultra thick] (6,0.5) -- (6,-0.5);
\draw[ultra thick]  (0, 0) edge[<->] node[midway,fill=white] {epoch} (2, 0);
\draw[ultra thick]  (2, 0) edge[<->] node[midway,fill=white] {epoch} (4, 0);
\draw[ultra thick]  (4, 0) edge[<->] node[midway,fill=white] {epoch} (6, 0);
\draw[decorate sep={1mm}{4mm},fill] (6.2,0) -- (7.9,0);
\draw[ultra thick] (8,0.5) -- (8,-0.5);
\draw[ultra thick] (10,0.5) -- (10,-0.5);
\draw[ultra thick] (12,0.5) -- (12,-0.5);
\draw[ultra thick] (14,0.5) -- (14,-0.5);
\draw[ultra thick]  (8, 0) edge[<->] node[midway,fill=white] {epoch} (10, 0);
\draw[ultra thick]  (10, 0) edge[<->] node[midway,fill=white] {epoch} (12, 0);
\draw[ultra thick]  (12, 0) edge[<->] node[midway,fill=white] {epoch} (14, 0);
\draw[ultra thick]  (0, -1) edge[<->] node[midway,fill=white] {long interval} (14, -1);
\end{tikzpicture}
\caption{Long interval, divided
into epochs.}
\label{fig:scalesep}
\end{figure}
epoch we work out the empirical correlation matrix $C_{\mathrm{ep},i} ,
\ i=1,\ldots,N_{\mathrm{ep}}$. Hence, a smaller numerical value of the
index $i$ indicates an earlier, a larger numerical value a later
epoch. To illustrate the non-stationarity for the example of a
financial market, we consider $K=479$ stocks in the New York Stock
Exchange (NYSE) in the year 2014~\cite{NYSE_2014}. From this data, we derive as amplitudes the returns, \textit{i.e.} relative price changes, with a time resolution of one second and compute correlation matrices for the epochs. We divide the year 2014 into
$N_{\mathrm{ep}}=250$ epochs, \textit{i.e.}, into 250 intraday data
sets. We use a normalized Frobenius inner product of two matrices as
a pairwise similarity measure. As a reference, we employ the average
of the epoch correlation matrices
\begin{equation} \label{eqn:AvgCorrMat}
\overline{C} = \frac{1}{N_{\mathrm{ep}}} \sum_{i=1}^{N_{\mathrm{ep}}}  C_{\mathrm{ep},i}  \ .
\end{equation}
The normalized Frobenius inner product of an individual epoch correlation matrix
$C_{\mathrm{ep},i}$ and the average
\begin{equation} \label{eqn:CosineAlpha_3}
\cos \widetilde{\alpha}_{i} = \frac{\mathrm{tr\,}C_{\mathrm{ep},i} \overline{C}}
{ \sqrt{ \mathrm{tr\,} C_{\mathrm{ep},i}^2 \; \mathrm{tr\,} \overline{C}^{\,2} }} 
\end{equation}
defines the cosine of a generalized angle $\widetilde{\alpha}_{i}$. As
depicted in Fig.~\ref{fig:twocorr_4}, non--stationarity makes the
results fluctuate, but the relatively large values of $\cos
\widetilde{\alpha}_{i}$ indicate that the matrices $C_{\mathrm{ep},i}$
have a gross structure in common. In finance, it is given by the
industrial sectors. More generally, it means that we have to
incorporate this gross structure in our model. We come back to this
point.  To demonstrate how large the fluctuations about this gross
structure are, we also look at the mutual normalized inner products of
the residuals $C_{\mathrm{ep},i}-\overline{C}$, given by
\begin{equation} \label{eqn:CosineAlpha}
\cos \alpha_{ij} = \frac{\mathrm{tr\,}(C_{\mathrm{ep},i} - \overline{C}) (C_{\mathrm{ep},j} - \overline{C})}
{ \sqrt{\mathrm{tr\,} (C_{\mathrm{ep},i} - \overline{C})^2 \;
\mathrm{tr\,} (C_{\mathrm{ep},j} - \overline{C})^2 }} \ .
\end{equation}
In Fig.~\ref{fig:twocorr}, we represent the values $\cos \alpha_{ij}$
in matrix form. As seen, the correlation matrices exhibit stronger
non--stationarity at the beginning of 2014.  Towards the end of 2014,
the periods during which correlation matrices are similar become
longer.  Different quasi--stationary periods are clearly visible in
Fig.~\ref{fig:twocorr}.  These periods of greater similarity between
correlation matrices are related to the market states identified in
Ref.~\cite{munnixIdentifyingStatesFinancial2012,Chetalova_2015,Chetalova_2015_2,Stepanov_2015_MultiAsset,Stepanov_2015,Rinn_2015,Pharasi_2018,Heckens_2020,heckens2021new,heckens2022new,marti2021review,pharasi2020market,pharasi2021dynamics,James_2022,Wand_2023,Hessler_2023,Wand_2023_2}. The probability density function
of the $\cos \alpha_{ij}$, depicted in Fig.~\ref{fig:twocorr_2},
reflects a remarkable variation of the residuals. This fluctuation
about the gross structure is the one we capture with our model.

We mention in passing that the cosines $\cos \alpha_{ij}$ and $\cos
\widetilde{\alpha}_{i}$ are similar to but different from Pearson
correlation coefficients, as the normalization or referencing,
respectively, is not the same.

In our model, we divide the long time interval into short epochs on
which we assume, as discussed above, relatively small variations of
the correlations.  Conceptually, we may even drop this assumption,
although it provides a convenient guideline for a data analysis. All
what matters is that in our model we view the fluctuations of the
correlations on the long time interval as pieced together from the
individual epochs. Naturally, the multivariate distribution in the
epochs on the one hand and on the long time interval on the other hand
ought to be clearly different. In our model, developed in
Ref.~\cite{Schmitt_2013,Guhr_2021} and further extended in
Ref.~\cite{Guhr_2021}, we make the assumption that the multivariate
amplitude distributions in the different epochs have the same shape,
\textit{i.e.}~the same functional form, and differ only in the
measured correlation matrices $C_\mathrm{ep}$.  The challenge is to
choose a functional form for $p(r|C_{\mathrm{ep}})$ that properly fits
the data in all epochs such that the non--stationary variations are
captured by the correlation matrix $C_{\mathrm{ep}}$ which differs
from epoch to epoch. The set of correlation matrices $C_\mathrm{ep}$
measured in all epochs is a truly existing ensemble which we now model
by an ensemble of random correlation matrices $XX^\dagger/N$. The
model data matrices $X$ have dimension $K\times N$ where $N$ is the
length of the model time series. For each value of $N$, the matrices
$XX^\dagger/N$ have dimension $K\times K$, which allows us to use $N$
as a tunable parameter. As will become clearer later on, the larger
$N$, the smaller the fluctuations of the model correlations. We draw
$X$ from a random matrix distribution $w(X|C,D)$. Here, $C$ and $D$
are the sample correlation matrices for time and position series,
respectively, measured over the long time interval. To construct the
multivariate amplitude distribution on the long time interval, we
replace in $p(r|C_{\mathrm{ep}})$ the epoch correlation matrices by
the random ones,
\begin{equation} \label{eq:rep}
C_{\mathrm{ep}} \longrightarrow \frac{1}{N} X X^{\dagger} \ ,
\end{equation}
and integrate over the ensemble
\begin{equation} \label{eqn:RMTmodelStartingPoint}
\langle p \rangle(r|C,D) = \int p\left(r\bigg \vert \frac{1}{N} X X^\dagger\right) w(X|C,D)d[X] \ ,
\end{equation}
where the measure $d[X]$ is the product of the differentials of all independent
variables, see Ref.~\cite{Guhr_2021}. This
ensemble random matrix average is meant to capture a truly existing
matrix ensemble, namely that of the epoch correlation matrices
$C_{\mathrm{ep}}$, while most other random matrix models are based on the
concept of second ergodicity, \textit{i.e.}~to model statistical
features of one large spectrum, an average over a fictitious ensemble
of random matrices is employed. Hence, as our random matrix model does
not ground on second ergodicity, the dimension of the correlation
matrices considered does not have to be large either. Our model applies
to correlation matrices of any size. 

\subsection{\label{subsec:agg}Rotation and Aggregation Procedure}

To make data analyses feasible, we will restrict the choices for
$p(r|C_{\mathrm{ep}})$ and $w(X|C,D)$ such that $p(r|C_{\mathrm{ep}})$ and
$\langle p \rangle (r|C,D)$ depend on the amplitudes only via the squared
Mahalanobis distances \cite{MahalanobisReprint2018} $r^{\dagger}
C^{-1}_\mathrm{ep} r$ and $r^{\dagger} C^{-1} r$, respectively.
The diagonalization of the correlation matrix $C$ reads
\begin{eqnarray}
C=U \Lambda U^{\dagger} \qquad \mathrm{with} \qquad \Lambda = \mathrm{diag}(\Lambda_1, \ldots, \Lambda_K) \ , \quad \Lambda_k>0 
\end{eqnarray}
with an an orthogonal matrix $U$. The same applies to
$C_{\mathrm{ep}}$ with eigenvalues $\Lambda_{\mathrm{ep},k}$. For the
inverse correlation matrices, we then have $C^{-1} = U \Lambda^{-1}
U^{\dagger}$. In the data analyses, we will always work with
full--rank correlation matrices which warrants the existence of their
inverse. For the squared Mahalanobis distance, we find
\begin{eqnarray}
r^{\dagger} C^{-1} r = r^{\dagger} U \Lambda^{-1} U^{\dagger} r = \bar{r}^\dagger \Lambda^{-1} \bar{r}
= \sum_{k=1}^K \frac{\bar{r}_k^2}{\Lambda_k}
\end{eqnarray}
with
\begin{eqnarray}\label{rescaled}
\bar{r} = U^\dagger r \ , \qquad \bar{r}_k = \sum_{l=1}^K  U_{lk} r_l
\end{eqnarray}
and similarly for $C_{\mathrm{ep}}$. Thus, going from the amplitudes $r$
to the linear combinations $\bar{r}$, we rotate into the eigenbasis of
the correlation matrix. If the covariance matrices are used instead of
the correlation matrices, everything works in the same way
\textit{mutatis mutandis}.  Integrating out all variables in $\bar{r}$
but one $\bar{r}_k$, say, we obtain $K$ univariate amplitude distributions
\begin{eqnarray}
\hspace{0.55cm}p^{(\mathrm{rot},k)} (\bar{r}_k|C_{\mathrm{ep}}) &= \int p(\bar{r}|C_\mathrm{ep}) d [ \bar{r} ]_{\neq k}  \nonumber\\
\langle p\rangle^{(\mathrm{rot},k)} (\bar{r}_k|C,D) &= \int \langle p\rangle (\bar{r}|C,D) d [ \bar{r} ]_{\neq k}
\label{multiuni}
\end{eqnarray}
for each epoch and for the longer interval, respectively. The
integrals are best carried out by inserting the characteristic
functions. The distributions
$p^{(\mathrm{rot},k)} (\bar{r}_k|C_{\mathrm{ep}})$ have the same
functional form for all $k$ in each epoch and, similarly, the
distributions $\langle p\rangle^{(\mathrm{rot},k)} (\bar{r}_k|C,D)$ for
all $k$ on the long interval. However, their parameters, more precisely, the
eigenvalues entering, are different. These $K$ distributions provide full
information on the correlated multivariate system, because all linear
combinations differ. 

The calculation also reveals that the resulting univariate
distributions for the rotated amplitudes have the same functional form
as the univariate distributions $p^{(\mathrm{orig},k)}
(r_k|C_{\mathrm{ep}})$ and $\langle p \rangle^{(\mathrm{orig},k)}
(r_k|C,D)$ for the unrotated, original amplitudes, but of course with
changes in the parametrical dependence, see
Secs.~\ref{sec:RMTAmplitudeDist} and \ref{sec:RMTEnsembleDist}.

Anticipating the forthcoming data analysis, we briefly sketch the
procedure of aggregation.  To accumulate data for statistical
significance, we normalize the $\bar{r}_k$ to the square root of the
corresponding eigenvalue,
\begin{equation}\label{eq:widetilde}
\widetilde{r}_k = \frac{\bar{r}_k}{\sqrt{\Lambda_{k}}}  ,
\end{equation}
or $\Lambda_{\mathrm{ep},k}$, respectively, and lump together all $K$
distributions. This yields statistically highly significant univariate
empirical distributions $p^{(\mathrm{aggr})} (\widetilde{r})$ and
$\langle p\rangle^{(\mathrm{aggr})} (\widetilde{r})$ which
facilitates a careful study of the tail behavior.

\subsection{\label{sec:RMTAmplitudeDist}Choice of Multivariate Amplitude Distributions in the Epochs}

We choose two functional forms for the amplitude distributions in the
epochs. We make the assumption that non--Markovian effects may be
neglected on shorter time scales, \textit{i.e.}~in the epochs. Their
inclusion would be mathematical feasible, but would lead to much more
complicated formulae.  In most complex systems that we have worked
with, a wise choice of the epoch length can always justify the
neglection of non--Markovian effects within the epochs.  We notice that
short--term memory effects are known to exist in correlated financial
markets~\cite{WSG2015preprint,wang2016cross,wang2016average,benzaquen2017dissecting},
but they are small.  A first choice for the multivariate amplitude distributions
is the Gaussian
\begin{equation}\label{eqn:MultiGaussian}
p_\mathrm{G}(r|C_{\mathrm{ep}})=\frac{1}{\sqrt{\mathrm{det}2 \pi C_{\mathrm{ep}}}}
\exp{ \left(-\frac{1}{2} r^{\dagger} C_{\mathrm{ep}}^{-1}   r  \right)}.
\end{equation}
The measured correlation matrix
\begin{equation}\label{eqn:corrG}
C_{\mathrm{ep}} = \langle rr^\dagger\rangle_{\mathrm{ep}} 
\end{equation}
differs from epoch to epoch. The rotated univariate amplitude
distributions are
\begin{equation} \label{eq:RotRescGaussian}
p_\mathrm{G}^{(\mathrm{rot},k)} (\bar{r}_k|\Lambda_{\mathrm{ep},k}) = \frac{1}{\sqrt{2 \pi\Lambda_{\mathrm{ep},k}}}
\exp \left( - \frac{\bar{r}_k^2}{2\Lambda_{\mathrm{ep},k}} \right) .
\end{equation}
If the covariance matrices $\Sigma_{\mathrm{ep}}$ instead of the correlation matrices $C_{\mathrm{ep}}$ are used,
the $\Lambda_{\mathrm{ep},k}$ are the eigenvalues of the former.

The univariate distributions of the original (unrotated) amplitudes
can also be heavy-tailed for various reasons, see finance~\cite{Cont_2001,Gabaix2003,Farmer_2004,Farmer_2004_2,Schmitt_2012} as an example. This
prompts our second choice \cite{Guhr_2021}
\begin{eqnarray}\label{eq:DetVer}
p_\mathrm{A}(r|\hat{C}_{\mathrm{ep}}) = \sqrt{\frac{2}{m}}^K \frac{\Gamma(l)}{\Gamma(l-K/2)}
\frac{1}{\sqrt{\det 2\pi \hat{C}_\mathrm{ep}}}
\frac{1}{\displaystyle\left(1+\frac{1}{m} r^\dagger \hat{C}^{-1}_\mathrm{ep} r\right)^l} 
\end{eqnarray}
with an algebraic tail determined by the power $l$. Here and in the
sequel, the indices $\mathrm{G}$ and $\mathrm{A}$ stand for Gaussian
or algebraic shape of the distributions, respectively.  We notice that
the input matrix $\hat{C}_\mathrm{ep}$ which is to be measured for
each epoch can in the algebraic case not directly be identified with
the sample correlation or covariance matrix. However, the simple
relation \cite{Guhr_2021}
\begin{eqnarray}
C_\mathrm{ep} = \langle rr^\dagger\rangle_\mathrm{ep} = \beta_\mathrm{A} \hat{C}_\mathrm{ep}
\label{eq:GauVerM1}
\end{eqnarray}
with
\begin{eqnarray}
\beta_\mathrm{Y}  = \cases{
\displaystyle
1               \ ,      &  if  \ Y=G  \\
\displaystyle
\frac{m}{2l-K-2} \ ,      & if  \ Y=A
} \ .
\label{eq:GauVerR1beta}
\end{eqnarray}
holds for the expectation value $\langle rr^\dagger\rangle$ as an
estimator for the sample correlations or covariances.

Importantly, the relation (\ref{eq:GauVerM1}) allows us to fix one of
the two parameters $l$ or $m$ in this algebraic case $\mathrm{Y=A}$. We choose
the latter and replace $m\hat{C}_\mathrm{ep}$ with $(2l-K-2)C_\mathrm{ep}$ such that
\begin{eqnarray}\label{eq:DetVer2}
p_\mathrm{A}(r|C_\mathrm{ep}) &= \sqrt{\frac{2}{2l-K-2}}^K \frac{\Gamma(l)}{{\Gamma(l-K/2)}}
\nonumber\\
&\qquad \frac{1}{\sqrt{\det 2\pi C_\mathrm{ep }}} \frac{1}{\displaystyle\left(1+\frac{1}{2l-K-2} r^\dagger C_\mathrm{ep}^{-1} r\right)^l} \ .
\end{eqnarray}
In this multivariate distribution, $l$ is the only fit parameter. 

Viewed as function of the Mahalanobis distance
$\sqrt{r^{\dagger}\hat{C}^{-1}_{\mathrm{ep}} r}$, the distribution
(\ref{eq:DetVer}) is of a generalized Student $t$ type
\cite{theodossiou1998financial}. In the standard univariate Student $t$ distribution
with $m$ degrees of freedom, one has $l=(m+1)/2$. Multivariate $t$
distributions were considered for financial data in
Refs.~\cite{CTJ1988,VPW1989,OS1993} and particularly in
Ref.~\cite{KanZhou2017}. In such $K$ multivariate $t$ distributions,
with $m$ degrees of freedom, the relation $l=(m+K)/2$ holds.  In our
choice, the two parameters $l$ and $m$ are first independent and then related (\ref{eq:GauVerM1}), (\ref{eq:GauVerR1beta}). The $K$
multivariate distribution (\ref{eq:DetVer}) is normalizable, if $l >
K/2$.

Analogously to Eq.~(\ref{eq:RotRescGaussian}), we calculate the
univariate distributions for the rotated amplitudes and find, not
surprisingly, a formula with a reduced power
\begin{eqnarray}\label{eq:RotRescAlgebraic}
p_\mathrm{A}^{(\mathrm{rot},k)} (\bar{r}_k|\Lambda_{\mathrm{ep},k}) &=
\frac{1}{\sqrt{\pi(2l_{\mathrm{rot}}-3)\Lambda_{\mathrm{ep},k}}} \frac{\Gamma(l_{\mathrm{rot}})}{\Gamma(l_{\mathrm{rot}}-1/2)} \\\nonumber
&\qquad
\frac{1}{\displaystyle\left(1 + \frac{\bar{r}_k^2}{(2l_{\mathrm{rot}}-3)\Lambda_{\mathrm{ep},k}}\right)^{l_{\mathrm{rot}}}} \ .
\end{eqnarray}
The combined parameter
\begin{equation}\label{eq:laggr}
l_{\mathrm{rot}} = l -\frac{K-1}{2}
\end{equation}
occurs because we integrated out $K-1$ variables of the $K$
multivariate distribution. We notice that the $\Lambda_{\mathrm{ep},k}$
are the eigenvalues of the sample correlation matrix.

Importantly, the corresponding univariate distributions
$p_{\mathrm{Y}}^{(\mathrm{orig},k)} (r_k|C_{\mathrm{ep}})$ for the original, unrotated
amplitudes have the same functional forms. They follow from
Eqs.~(\ref{eq:RotRescGaussian}) and (\ref{eq:RotRescAlgebraic}) by simply replacing
$\Lambda_{\mathrm{ep},k}$ with the number one or with the variances
$\Sigma_{\mathrm{ep},kk} = \sigma_{\mathrm{ep},k}^2$ if correlation or
covariance matrices are used, respectively.  The information on the
multivariate, correlated system is contained in the eigenvalues of the
correlation or covariance matrices which appear explicitly in the
univariate distributions of the rotated amplitudes, but not in those of
the original, unrotated amplitudes. Hence the latter do not carry
information on the multivariate, correlated system, in contrast to the
former.

\subsection{\label{sec:RMTEnsembleDist}Choice of Ensembles for the Fluctuating Correlations}

To capture the non--stationarity, we model the fluctuations of
correlations by random $K\times N$ data matrices $X$ which we draw
from either Gaussian or algebraic distributions.  
Our
first choice for the ensemble distribution is the multimultivariate
Gaussian
\begin{equation}\label{eqn:wG}
w_\mathrm{G}(X|C,D) = \frac{1}{\sqrt{\mathrm{det} 2 \pi D \otimes C}} \exp{ \left( -\frac{1}{2}\mathrm{tr}D^{-1}X^\dagger C^{-1}X \right)} \ .
\end{equation}
In statistical inference, this is the celebrated doubly correlated
Wishart distribution with input matrices $C$ and $D$ describing the
correlation structure of the time and position series, $r(t)$ and
$\widetilde{r}_k$, respectively. They can be determined by sampling over the
long time interval,
\begin{eqnarray}
C = \langle rr^\dagger\rangle \qquad \mathrm{and} \qquad D = \langle \widetilde{r}\widetilde{r}^\dagger\rangle \ ,
\label{eq:CD}
\end{eqnarray}
where $D$ measures the memory effects. 
In our model these are the non-Markovian effects across epochs.
The random data matrix in the
model has dimension $K\times N$, thus the time series have length $N$
which is an adjustable parameter controlling how strongly the $K\times
K$ model correlation matrix $XX^\dagger/N$ and the $K\times K$ model
correlation matrix $X^\dagger X/K$ fluctuate about the input matrices
$C$ and $D$, respectively.  Thus, it is a feature of our model that
the length $N$ of the random model time series is different from, in
general much shorter than, the length $T$ of the long
interval. Although the input matrix $D$ in Eq.~(\ref{eqn:wG}) and the
sampled correlation matrix $D$ in (\ref{eq:CD}) are different due to
the structure of our model, we do not distinguish them in the notation.
From a practical point of view, the input matrix $D$ should contain
the sector of the sampled matrix $D$ corresponding to the largest
eigenvalues.

As explained in Sec.~\ref{sec:IdeaAndConcept}, our model is quite
different from statistical inference, second ergodicity is not evoked
as the random ensemble (\ref{eqn:wG}) models the truly existing ensemble of
the measured epoch correlation matrices $C_\mathrm{ep}$. Hence, the
$K\times K$ model correlation matrices $XX^\dagger/N$ do not have to
be large.

To model possible heavy tails in the fluctuations of the correlations,
we also choose the multimultivariate, algebraic determinant
distribution
\begin{eqnarray}\label{eqn:wA}
w_\mathrm{A}(X|\hat{C},\hat{D}) &= \left(\frac{2}{M}\right)^{\frac{K N}{2}} \prod_{n = 1}^{N}
\frac{\Gamma(L-(n-1)/2)}{\Gamma(L-(K+n-1)/2)}\frac{1}{\sqrt{\mathrm{det} 2 \pi \hat{D} \otimes \hat{C}}} \nonumber\\
&\qquad\qquad\qquad\qquad \frac{1}{\displaystyle\mathrm{det}^L\left(\mathbb{1}_{N} + \frac{1}{M} \hat{D}^{-1}X^\dagger \hat{C}^{-1}X \right)} 
\end{eqnarray}
depending on two input parameter correlation matrices $\hat{C}$ and
$\hat{D}$.  It generalizes the matrixvariate $t$ distribution with
$\nu$ degrees of freedom \cite{GuptaNagar2000} which is recovered for
$M=1$ and $L=(\nu+K+N-1)/2$. Once more, the indices $\mathrm{G}$ and
$\mathrm{A}$ stand for Gaussian or algebraic shape of the
distributions, respectively.  In our application, $L,M,N$ in the
algebraic case are first independent parameters, but the relations
\cite{Guhr_2021}
\begin{eqnarray}
C = B_\mathrm{A} \hat{C} \qquad \mathrm{and} \qquad D = B_\mathrm{A} \hat{D}
\label{eq:CDrel}
\end{eqnarray}
between the sample and the input parameter correlation matrices with
\begin{eqnarray}
B_\mathrm{Y'}  = \cases{
\displaystyle
1                 \ ,      & if  \ $\mathrm{Y^\prime}$=G  \\
\displaystyle
\frac{M}{2L-K-N-1}\ ,      & if  \ $\mathrm{Y^\prime}$=A
} \ .
\label{eq:GauVerR1Beta}
\end{eqnarray}
facilitate the elimination of the parameter $M$. In Eq.~(\ref{eqn:wA}), we replace $M\hat{C}$ with $(2L-K-N-1)C$ and find
\begin{eqnarray}\label{eqn:wAM}
w_\mathrm{A}(X|C,D) &= \left(\frac{2}{2L-K-N-1}\right)^{\frac{K N}{2}}
\nonumber\\
&\qquad
\prod_{n = 1}^{N}
\frac{\Gamma(L-(n-1)/2)}{\Gamma(L-(K+n-1)/2)}\frac{1}{\sqrt{\mathrm{det} 2 \pi D \otimes C}} \nonumber\\
&\qquad\qquad \frac{1}{\displaystyle\mathrm{det}^L\left(\mathbb{1}_{N} + \frac{1}{2L-K-N-1} D^{-1}X^\dagger C^{-1}X \right)} \ .
\end{eqnarray}
Importantly, as the dependence of $w_\mathrm{A}(X|\hat{C},\hat{D})$ on
the input matrices $\hat{C}$ and $\hat{D}$ is, apart from their
dimensions, fully symmetric, the replacements of $M\hat{C}$ with
$(2L-K-N-1)C$ and of $M\hat{D}$ with $(2L-K-N-1)D$ are equivalent.
Hence the ensemble averages $\langle XX^\dagger/N\rangle$ and
$\langle X^\dagger X/K\rangle$ now yield $C$ and $D$ as required.

\subsection{\label{sec:RMTEnsembleAverageDist}Resulting Multivariate Amplitude Distributions on the Long Interval}

Employing Eq.~(\ref{eqn:RMTmodelStartingPoint}), we calculate the
multivariate amplitude distributions $\langle p \rangle_{YY'}(r|C,D)$ on
the long interval. As the multivariate amplitude distributions
$p_\mathrm{Y}(r|C_\mathrm{ep})$ and the ensemble distributions
$w_\mathrm{Y'}(X|C,D)$ both come in a Gaussian $\mathrm{Y=G}$ and an algebraic
$\mathrm{Y=A}$ choice according to
Eqs.~(\ref{eqn:MultiGaussian}), (\ref{eq:DetVer}), (\ref{eqn:wG}), (\ref{eqn:wA}),
we arrive at four distributions $\langle p \rangle_{YY'}(r|C,D)$ on
the long interval. Details of the calculations can be found in
Ref.~\cite{Guhr_2021}, we only present the results. Remarkably, almost
all integrals can be done, the formulae are fairly compact, given the
complexity of the model. It is an important feature of our model, that
all multivariate distribution on the long interval depend on the
amplitudes only via the Mahalanobis distances
\cite{MahalanobisReprint2018} $\sqrt{r^{\dagger} C^{-1}
r}$. Explicitly, our results are
\begin{eqnarray} \label{eqn:PGG}
\langle p \rangle_{\mathrm{GG}}(r|C,D) &= \frac{1}{\sqrt{r^\dagger C^{-1} r}^{(K-2)/2}} \frac{1}{\sqrt{\det 2\pi C}}
\nonumber\\
&\qquad
\int\limits_0^{\infty}\frac{J_{(K-2)/2}\left(\rho\sqrt{r^\dagger C^{-1} r}\right)}{\sqrt{\det(\mathbb{1}_N + D \rho^2/N)}}\rho^{K/2}d\rho  
\end{eqnarray}
in the Gaussian--Gaussian case,
\begin{eqnarray} \label{eqn:PGA}
\langle p \rangle_{\mathrm{GA}}(r|C,D) &= \frac{\Gamma(L-(N-1)/2)}{\Gamma(K/2)\Gamma(L-(K+N-1)/2)}
\nonumber\\
&\qquad
\frac{1}{\sqrt{\det 2\pi C (2L-K-N-1)/N}}
\nonumber\\
& \qquad\qquad \int\limits_0^{\infty} {_1F_1}\bigg(L-\frac{N-1}{2},\frac{K}{2},
\nonumber\\
& \qquad\qquad\qquad
-\frac{u N}{2(2L-K-N-1)} r^\dagger C^{-1} r \bigg) \nonumber\\
& \qquad\qquad\qquad\qquad \frac{u^{K/2-1} du}{\sqrt{\det(\mathbb{1}_N + u D)}} 
\end{eqnarray}
in the Gaussian-Algebraic case,
\begin{eqnarray} \label{eqn:PAG}
\langle p \rangle_{\mathrm{AG}}(r|C,D) &= \frac{\Gamma(l)}{\Gamma(K/2)\Gamma(l-K/2)} \frac{1}{\sqrt{\det 2\pi C (2l-K-2)/N}}\nonumber\\
& \qquad  \int\limits_0^{\infty}{_1F_1}\left(l,\frac{K}{2},-\frac{u N}{2(2l-K-2)} r^\dagger C^{-1} r \right) \nonumber\\                                         
\hspace{-2.5cm}                     & \qquad\qquad  \frac{u^{K/2-1} du}{\sqrt{\det(\mathbb{1}_N + u D)}}
\end{eqnarray}
in the Algebraic--Gaussian case and finally
\begin{eqnarray} \label{eqn:PAA}
\langle p \rangle_{\mathrm{AA}}(r|C,D) &= \frac{\Gamma(l)\Gamma(l-(N-1)/2)}{\Gamma(K/2)\Gamma(l-K/2)\Gamma(L-(K+N-1)/2)}\nonumber\\
& \qquad  \frac{1}{\sqrt{\det \pi C (2l-K-2)(2L-K-N-1)/N}}\nonumber\\
& \qquad \int\limits_0^{\infty}{_2F_1}\bigg(l,L-\frac{N-1}{2},\frac{K}{2},
\nonumber\\
& \qquad\qquad
-\frac{u N}{(2l-K-2)(2L-K-N-1)} r^\dagger C^{-1} r\bigg)\nonumber\\
& \qquad\qquad\qquad \frac{u^{K/2-1} du}{\sqrt{\det(\mathbb{1}_N + u D)}}
\end{eqnarray}
in the Algebraic--Algebraic case.  For the occurring special functions
Bessel $J_{\nu}$, Macdonald $K_{\nu}$, Kummer $_1F_1$, Tricomi $U$ and
hypergeometric Gauss $_2F_1$ we use the definitions and conventions of
Ref.~\cite{NIST:DLMF}. These multivariate distributions $\langle p
\rangle_{\mathrm{YY'}}(r|C,D)$ still include non--Markovian effects
encoded in the input correlation matrix $D$ of the position series. We
notice that only its eigenvalues enter the multivariate
distributions.
A thorough study of memory effects in the context of our model should
be carried out in systems such as climate or traffic where their role
can be clearly distinguished.
The Markovian case $D=\mathbb{1}_N$ is of particular interest.

We derive the univariate distributions of the rotated amplitudes
$\bar{r}_k$, calculate the integrals over the
other $K-1$ rotated amplitudes, define the combined parameter
\begin{equation}\label{eq:Laggr}
L_{\mathrm{{rot}}} = L -\frac{K-1}{2}
\end{equation}
analogously to $l_{\mathrm{rot}}$ in Eq.~(\ref{eq:laggr}) and arrive at
\begin{eqnarray} \label{eqn:RotRescPGG}
\langle p \rangle_{\mathrm{GG}}^{(\mathrm{rot},k)} (\bar{r}_k|\Lambda_k) &=
\frac{1}{2^{(N-1)/2}\Gamma (N/2)}\sqrt{\frac{N}{\pi\Lambda_k}}
\left(\frac{N\bar{r}^2}{\Lambda_k}\right)^{(N-1)/4}
\nonumber\\
& \qquad
K_{(1-N)/2}\left(\sqrt{\frac{N\bar{r}^2}{\Lambda_k}}\right) 
\end{eqnarray}
in the Gaussian--Gaussian case,
\begin{eqnarray} \label{eqn:RotRescPGA_New}
\langle p \rangle_{\mathrm{GA}}^{(\mathrm{rot},k)}(\bar{r}_k|\Lambda_k) &= \frac{\Gamma(L_{\mathrm{rot}}-(N-1)/2)\Gamma(L_{\mathrm{rot}})}
{\Gamma(L_{\mathrm{rot}}-N/2)\Gamma(N/2)}
\nonumber\\
& \qquad
\sqrt{\frac{N}{2\pi(2L_{\mathrm{rot}}-N-2)\Lambda_k}}  U\bigg(L_{\mathrm{rot}}-\frac{N-1}{2},
\nonumber\\
& \qquad\qquad\qquad
\frac{1-N}{2}+1,\frac{N\bar{r}^2}{2(2L_{\mathrm{rot}}-N-2)\Lambda_k}\bigg) 
\end{eqnarray}
in the Gaussian--Algebraic case,
\begin{eqnarray} \label{eqn:RotRescPAG_New}
\langle p \rangle_{\mathrm{AG}}^{(\mathrm{rot},k)}(\bar{r}_k|\Lambda_k) &= \frac{\Gamma(l_{\mathrm{rot}})\Gamma(l_{\mathrm{rot}}+(N-1)/2)}
{\Gamma(l_{\mathrm{rot}}-1/2)\Gamma(N/2)}
\nonumber\\
& \qquad
\sqrt{\frac{N}{2\pi(2l_{\mathrm{rot}}-3)\Lambda_k}} \nonumber\\
&\qquad\qquad U\left(l_{\mathrm{rot}},\frac{1-N}{2}+1,\frac{N\bar{r}_k^2}{2(2l_{\mathrm{rot}}-3)\Lambda_k}\right) 
\end{eqnarray}
in the Algebraic--Gaussian case and finally
\begin{eqnarray} \label{eqn:RotRescPAA_New}
\langle p \rangle_{\mathrm{AA}}^{(\mathrm{rot},k)}(\bar{r}_k|\Lambda_k) &=
\frac{\Gamma(l_{\mathrm{rot}})\Gamma(l_{\mathrm{rot}}+(N-1)/2)}
{\Gamma(l_{\mathrm{rot}}-1/2)\Gamma(L_{\mathrm{rot}} + l_{\mathrm{rot}})} \nonumber\\
&\qquad\frac{\Gamma(L_{\mathrm{rot}})\Gamma(L_{\mathrm{rot} } -(N-1)/2)}
{\Gamma(L_{\mathrm{rot}}-N/2)\Gamma(N/2)} \nonumber\\
&\qquad\qquad \sqrt{\frac{N}{\pi(2L_{\mathrm{rot}}-N-2)(2l_{\mathrm{rot}}-3)\Lambda_k}} \nonumber\\
&\qquad\qquad _2F_1\bigg(l_{\mathrm{rot}},L_{\mathrm{rot}}- \frac{N-1}{2},L_{\mathrm{rot}}+l_{\mathrm{rot}},\nonumber\\
&\qquad\qquad\qquad      1-\frac{N\bar{r}_k^2}{(2L_{\mathrm{rot}}-N-2)(2l_{\mathrm{rot}}-3)\Lambda_k}\bigg) 
\end{eqnarray}
in the Algebraic--Algebraic case. The same remark as for univariate
distributions on the epochs applies.  The corresponding univariate
distributions $\langle
p\rangle_{\mathrm{YY'}}^{(\mathrm{orig},k)}(r_k|\Lambda_k)$ for the
original, unrotated amplitudes $r_k$ have the same functional form. They
follow from the above formulae by simply replacing $\Lambda_{k}$ with
the number one or with the variances $\Sigma_{kk} = \sigma_{k}^2$ if
correlation or covariance matrices are used,
respectively. Importantly, the equality of the functional forms for
the univariate distributions of original and rotated amplitudes does not
mean that the latter ones do not carry new information on the
multivariate system. The opposite is true. This information is encoded
in the eigenvalues of the correlation or covariance matrices which
enter the univariate distributions of the rotated amplitudes. Information
on the multivariate system can never be retrieved from only knowing
the univariate distributions of the original amplitudes.

Which are the parameters to be fitted in the above given univariate
distributions on the long interval? Of course, the number $K$ of
stocks, the sample correlation matrix $C$ and its eigenvalues
$\Lambda_k$ are known. The parameter $l$ or, equivalently,
$l_{\mathrm{rot}}$ has been determined by the fits of the epoch
distributions. Hence, for all distributions on the long interval, $N$
is a fit parameter and in the Gaussian--Algebraic and the
Algebraic--Algebraic cases, $L$ or, equivalently, $L_{\mathrm{rot}}$ is
a second fit parameter.
To carry out the fits in an unambiguous way, 
the parameter reduction as described in Sec.~\ref{sec:RMTEnsembleDist} 
is essential. For the reader with little experience in data analysis, 
we discuss this in App.~\ref{app:FitParameterReduction}.

We notice that due to our construction, the variances of the
univariate distributions for the rotated amplitudes are given
\cite{Guhr_2021} by
\begin{eqnarray}
\langle \bar{r}_k^2 \rangle_{\mathrm{YY'}}^{(\mathrm{rot},k)} = \Lambda_k 
\label{sec3.9a}
\end{eqnarray}
in all four cases $\mathrm{Y,Y'=G,A}$, where $\Lambda_k$ is eigenvalue of the
sample correlation or covariance matrix.

\subsection{\label{sec:GeneralLinearComb} Multivariate Distributions of Arbitrary Linear Combinations}

After constructing our model, the parameter $N$ and, if necessary, the
parameters $l$ and $L$ have to be determined. Due to the multivariate
character of the problem, this is a demanding task which we will carry
out for the example of financial data in II. Here, we want to give an
example for applications of our multivariate distributions on the long
interval, once the parameters have been determined. Consider $I$
functions $f_i(r), i=1,\ldots,I$ of the amplitudes $r$, ordered in an
$I$ component vector $f(r)$. The multivariate distributions of the $I$
variables $s_i, i=1,\ldots,I$, ordered in an $I$ component vector
$s$, which measure the thereby defined combinations of the amplitudes, follow from the $K$--dimensional filter integral
\begin{eqnarray}
\langle p \rangle_\mathrm{YY'}^{(\mathrm{comb})}(s \, \vert \, C,D,f) =
\int  \delta(s - f(r)) \langle p \rangle_{YY^\prime}(r\,\vert\, C,D) d [r] \ ,
\label{glc1}
\end{eqnarray}
where $\delta(s - f(r))$ is the product of the $I$ univariate
$\delta(s_i-f_i(r))$. We suppress the time dependence of the amplitudes
in our notation. Of particular interest for most aspects of risk management
are linear combinations,
\begin{eqnarray}
f_i(r) = \sum_{k=1}^K v_{ki} r_k = v_i^\dagger r \ , \quad \mathrm{where} \quad i=1,\ldots,I 
\label{glc2}
\end{eqnarray}
with constant coefficients $v_{ki}$.  It is useful to introduce the,
in general rectangular, $K \times I$ matrix containing the coefficient
vectors $v_i$ in the form
\begin{eqnarray}
V = [\begin{array}{@{}ccc@{}} v_1 & \cdots & v_{I} \end{array}]
= \left[\begin{array}{@{}cccc@{}} v_{11} & v_{12} & \cdots & v_{1I} \cr
\vdots & \vdots & & \vdots \cr
v_{K1} & v_{K2} & \cdots & v_{KI}\end{array}\right] \ .
\label{glc3}
\end{eqnarray}
In the context of financial returns, the vectors $v_i$ could for
instance correspond to certain indices, which weigh the $K$ assets
accordingly, or to any other choice in a portfolio selection. Using
our model which inherently accounts for non--stationarity, we are able
to calculate the multivariate portfolio return distribution of a
collection of $I$ such indices.

Given a coefficient matrix $V$, we now show how to calculate the
multivariate distribution of the $I$ variables $s$. We employ the
characteristic functions $\langle \varphi \rangle_\mathrm{YY'}(\omega\,
\vert\, C,D)$ as given in Ref.~\cite{Guhr_2021} for the multivariate
distributions in Eqs.~(\ref{eqn:PGG}) to (\ref{eqn:PAA}). They
contain the correlation matrix $C$ only via the bilinear product
$\omega^\dagger C \omega$. With an $I$--dimensional vector $\xi$ as
integration variable, we obtain
\begin{eqnarray}
\langle p \rangle_\mathrm{YY'}^{(\mathrm{comb})}(s \, \vert \, C,D,V) &= \int  \delta(s - V^\dagger r) \langle p \rangle_{YY^\prime}(r\,\vert\, C,D) d [r] \nonumber\\
&= \frac{1}{(2\pi)^{I}} \int  \int  e^{-i\xi^\dagger(s-V^\dagger r)} \langle p \rangle_{YY^\prime}(r\,\vert\, C,D) d [r] d [\xi]\nonumber\\
&= \frac{1}{(2\pi)^{I}} \int e^{-i\xi^\dagger s} \langle \varphi \rangle_{YY^\prime}\left(V\xi\,\vert\, C,D\right) d [\xi]\, .
\label{glc3}
\end{eqnarray}
This is an $I$--dimensional Fourier backtransform of the characteristic
function calculated in $K$--dimensional Fourier space. We notice that
$V^\dagger r$ is $I$--dimensional, whereas $V\xi$ is $K$--dimensional.
To make use of the fact that $\langle \varphi
\rangle_{YY^\prime}\left(V\xi\,\vert\, C,D\right)$ depends on the
bilinear product, we define the matrix
\begin{eqnarray}
\tilde C = V^\dagger C V = \left[\begin{array}{@{}cccc@{}}
v_1^\dagger Cv_1 & v_1^\dagger Cv_2 & \cdots & v_1^\dagger Cv_I \cr
\vdots & \vdots & & \vdots \cr
v_I^\dagger Cv_1 & v_I^\dagger Cv_2 & \cdots & v_I^\dagger Cv_I \end{array}\right] \ ,
\label{glc4}    
\end{eqnarray}
and further denote $\langle p
\rangle_{\mathrm{YY'}}^{(\mathrm{comb})}(s\,\vert\,C,D,V)$ as $\langle
p \rangle_{\mathrm{YY'}}^{(\mathrm{comb})}(s\,\vert\,\tilde C,D)$. We
notice that $\tilde C$ is not a correlation matrix, but as it is
positive semidefinite, it can be viewed as a covariance matrix. Of
course, upon rescaling $\tilde C$ may become a proper correlation
matrix.
As in Ref.~\cite{Guhr_2021} we change variables according to $\xi
\rightarrow \tilde C^{1/2}\xi$ and reduce the $I$--dimensional
Fourier-transform to a one--dimensional Hankel transform. If $C$ is
positive definite, $\tilde C$ will also be positive definite if $V$
has full column rank, which allows the change of variables. This means
that $V$ cannot define a linear
combination that is exactly replicated or canceled out by a linear
combination of the others vectors. Backtransformation in analogy to
Ref.~\cite{Guhr_2021} yields structurally similar solutions as
previously shown in Eqs.~(\ref{eqn:PGG}) to (\ref{eqn:PAA}), but now
depending on the squared Mahalanobis distance of the linear
combinations with the $I \times I$ matrix $\tilde C$. In finance
$\tilde C$ can be
understood as a matrix measuring dependence between the portfolio
returns of $I$ portfolios, rather than all $K$ stocks. We also notice
that the parameters of the resulting distributions contain the
dimension $I$, incorporating the effect that $I$--variate distributions
arise from
$K$--variate characteristic functions. The solutions finally are
\begin{eqnarray}
\langle p \rangle_{\mathrm{GG}}^{(\mathrm{comb})}(s\,\vert\,\tilde C,D) &= \frac{1}{\sqrt{s^\dagger \tilde C^{-1} s}^{({I} - 2)/2}} \frac{1}{\sqrt{\det 2\pi \tilde C}}\nonumber\\
&\qquad \int\limits_0^{\infty}\frac{J_{({I}-2)/2}\left(\rho\sqrt{s^\dagger \tilde C^{-1} s}\right)}{\sqrt{\det(\mathbb{1}_N + D \rho^2/N)}}\rho^{{I}/2}d\rho  
\label{glc5}
\end{eqnarray}
in the Gaussian--Gaussian case,
\begin{eqnarray}
\langle p \rangle_{\mathrm{GA}}^{(\mathrm{comb})}(s\,\vert\,\tilde C,D) &= \frac{\Gamma({I}/2 - N/2 + L-(K-1)/2)}{\Gamma({I}/2)\Gamma(L-(K+N-1)/2)}
\nonumber\\
&\qquad
\frac{1}{\sqrt{\det 2\pi \tilde C (2L-K-N-1)/N}}
\nonumber\\
& \qquad\qquad \int\limits_0^{\infty} {_1F_1}\bigg(\frac{{I}}{2} + L-\frac{K+N-1}{2},\frac{{I}}{2},
\nonumber\\
& \qquad\qquad\qquad
-\frac{u N}{2(2L-K-N-1)} s^\dagger \tilde C^{-1} s \bigg) \nonumber\\
& \qquad\qquad\qquad\qquad \frac{u^{{I}/2-1} du}{\sqrt{\det(\mathbb{1}_N + u D)}}
\label{glc6}
\end{eqnarray}
in the Gaussian--Algebraic case,
\begin{eqnarray}
\langle p \rangle_{\mathrm{AG}}^{(\mathrm{comb})}(s\,\vert\,\tilde C,D) &=
\frac{\Gamma(l-K/2 + {I}/2)}{\Gamma(l - K/2)\Gamma({I}/2)} \frac{1}{\sqrt{\det 2\pi \tilde C (2l-K-2)/N}}\nonumber\\
& \qquad  \int\limits_0^{\infty}{_1F_1}\left(l-\frac{K}{2}+\frac{{I}}{2},\frac{I}{2},-\frac{u N}{2(2l-K-2)} s^\dagger \tilde C^{-1} s \right) \nonumber\\                                         
\hspace{-2.5cm}                     & \qquad\qquad  \frac{u^{{I}/2-1} du}{\sqrt{\det(\mathbb{1}_N + u D)}}
\label{glc7}
\end{eqnarray}
in the Algebraic--Gaussian case,
\begin{eqnarray}
\langle p \rangle_{\mathrm{AA}}^{(\mathrm{comb})}(s\,\vert\,\tilde C,D) &=
\frac{\Gamma(l+{I}/2 - K/2)\Gamma({I}/2 + L -(K+N-1)/2)}{\Gamma({I}/2)\Gamma(l-K/2)\Gamma(L-(K+N-1)/2)}\nonumber\\
& \qquad  \frac{1}{\sqrt{\det \pi \tilde C (2l-K-2)(2L-K-N-1)/N}}\nonumber\\
& \qquad \int\limits_0^{\infty}{_2F_1}\bigg(l+\frac{{I}}{2} - \frac{K}{2},\frac{{I}}{2} - \frac{N}{2} + L-\frac{K-1}{2},\frac{{I}}{2},
\nonumber\\
& \qquad\qquad
-\frac{u N}{(2l-K-2)(2L-K-N-1)} s^\dagger \tilde C^{-1} s\bigg)\nonumber\\
& \qquad\qquad\qquad \frac{u^{{I}/2-1} du}{\sqrt{\det(\mathbb{1}_N + u D)}}
\label{glc8}
\end{eqnarray}
in the Algebraic--Algebraic case. For the choice $V=\mathbb{1}_K$ we
recover Eqs.~(\ref{eqn:PGG}) to (\ref{eqn:PAA}).

For Markovian dynamics, $D=\mathbb{1}_N$, the formulae simplify
further since the determinant containing $D$ reduces to an $N$--th
power. We arrive at
\begin{eqnarray}
\langle p \rangle_{\mathrm{GG}}^{(\mathrm{comb})}(s\,\vert\,\tilde C,\mathbb{1}_N) &= \frac{1}{2^{N/2-1}\Gamma(N/2)\sqrt{\det 2\pi \tilde C/N}}\nonumber\\
&\qquad \frac{K_{({I}-N)/2}(\sqrt{Ns^\dagger \tilde C^{-1}s})}{\sqrt{Ns^\dagger \tilde C^{-1}s}^{({I}-N)/2}}
\label{glc9}
\end{eqnarray}
in the Gaussian--Gaussian case,
\begin{eqnarray}
\langle p \rangle_{\mathrm{GA}}^{(\mathrm{comb})}(s\,\vert\,\tilde C,\mathbb{1}_N) &= \frac{\Gamma(L - (K-1)/2)}{\Gamma(N/2)\Gamma(L-(K+N-1)/2)} \nonumber\\
& \qquad \frac{\Gamma(L-(N-1)/2+{I}/2-K/2)}{\sqrt{\det 2\pi \tilde C (2L - K - N - 1)/N}} \nonumber\\
& \qquad U\bigg(\frac{{I}}{2}-\frac{K}{2}+L-\frac{N-1}{2}, \frac{{I}}{2}-\frac{N}{2}+1,\nonumber\\
& \qquad \qquad \frac{N}{2(2L-K-N-1)}s^\dagger \tilde C^{-1}s\bigg)
\label{glc10}
\end{eqnarray}
in the Gaussian--Algebraic case,
\begin{eqnarray}
\langle p \rangle_{\mathrm{AG}}^{(\mathrm{comb})}(s\,\vert\,\tilde C,\mathbb{1}_N) &= \frac{\Gamma(N/2-K/2+l)\Gamma(l+{I}/2-K/2)}{\Gamma(l-K/2)\Gamma(N/2)\sqrt{\det 2\pi \tilde C (2l - K - 2)/N}} \nonumber\\
& \qquad U\bigg( \frac{{I}}{2} - \frac{K}{2} + l, \frac{{I}}{2} - \frac{N}{2} + 1, \nonumber\\
& \qquad \qquad \frac{N}{2(2l - K - 2)}s^\dagger \tilde C^{-1} s\bigg)
\label{glc11}
\end{eqnarray}
in the Algebraic--Gaussian case, and finally
\begin{eqnarray}
\langle p \rangle_{\mathrm{AA}}^{(\mathrm{comb})}(s\,\vert\,\tilde C,\mathbb{1}_N) &= \frac{\Gamma({I}/2 - K/2 + l)\Gamma(l-(K-N)/2)}{\sqrt{\det\pi\tilde C (2l-K-2)(2L-K-N-1)/N} } \nonumber\\
& \qquad \frac{\Gamma({I}/2+L-(K+N-1)/2)}{\Gamma(l-K/2)\Gamma(L-(K+N-1)/2)\Gamma(N/2)} \nonumber\\
& \qquad \frac{\Gamma(L-(K-1)/2)}{\Gamma(L+l-K+({I}+1)/2)} \nonumber\\
&  \qquad {_2F_1}\bigg(l+\frac{{I}}{2}-\frac{K}{2}, \frac{{I}}{2}-\frac{K}{2} + L - \frac{N-1}{2}, \nonumber\\
&  \qquad \qquad L+l-K+\frac{{I}+1}{2}, \nonumber\\
&  \qquad \qquad 1 - \frac{N s^\dagger \tilde C^{-1} s}{(2l-K-2)(2L-K-N-1)}\bigg)
\label{glc12}
\end{eqnarray}
in the Algebraic--Algebraic case.

To facilitate the data comparison in the sequel, we look at special
cases of the above formulae. First, we set $I=1$ and provide the
univariate distribution of an arbitrary linear combination $s_1 =
v_1^\dagger r$, with a coefficient vector $v_1$, as a special case of
the above as
\begin{eqnarray}
\langle p \rangle_{\mathrm{YY'}}^{(\mathrm{comb}, 1)}(s_1 \,\vert\, v_1^\dagger C v_1)
= \frac{1}{2\pi} \int\limits_{-\infty}^{+\infty} e^{-i\xi_1 s_1} \langle \varphi \rangle_{\mathrm{YY'}} (\xi_1 v_1\,\vert\, C,D) d \xi_1 \ ,
\label{glc13}
\end{eqnarray}
where $\xi_1$ is now the one--dimensional Fourier variable.
In the Markovian case $D = \mathbb{1}_N$, we can carry out the calculation in analogy to Ref.~\cite{Guhr_2021}
and find
\begin{eqnarray}
\langle p \rangle_{\mathrm{YY'}}^{(\mathrm{comb}, 1)}(s_1 \, \vert \, v_1^\dagger C v_1)
= \langle p \rangle_\mathrm{YY'}^{(\mathrm{rot},1)}(s_1 \,\vert\, v_1^\dagger C v_1)\, ,
\label{glc14}
\end{eqnarray}
where the distributions $\langle p
\rangle_\mathrm{YY'}^{(\mathrm{rot},j)}$ are the univariate
distributions in Eqs.~(\ref{eqn:RotRescPGG}) to
(\ref{eqn:RotRescPAA_New}), but now the linear combination $s_1$ plays
the role of the univariate rotated return series $\bar r_k$ and the
bilinear product $v_1^\dagger C v_1$ plays the role of the
corresponding eigenvalue $\Lambda_k$. The parameters of the solutions
are exactly the same, regardless of whether they are calculated by
integrating out ${I}-1$ dimensions from $\langle p
\rangle_{\mathrm{AA}}^{(\mathrm{comb})}(s\,\vert\,\tilde C,D)$ or by
integrating out $K-1$ dimensions from $\langle
p\rangle_\mathrm{YY'}(r\, \vert\, C,D)$ as done in
Sec.~\ref{sec:RMTEnsembleAverageDist}.

In the bivariate setting $I = 2$, the distribution for the Algebraic--Algebraic case reads
\begin{eqnarray}
\langle p \rangle_{\mathrm{AA}}^{(\mathrm{comb})}(s_1, s_2\,\vert\,\tilde C,\mathbb{1}_N) &= \frac{\Gamma(1 - K/2 + l)\Gamma(l-(K-N)/2)}{\sqrt{\det\pi\tilde C (2l-K-2)(2L-K-N-1)/N} } \nonumber\\
& \qquad \frac{\Gamma(3/2+L-(K+N)/2)}{\Gamma(l-K/2)\Gamma(L-(K+N-1)/2)\Gamma(N/2)} \nonumber\\
& \qquad \frac{\Gamma(L-(K-1)/2)}{\Gamma(L+l-K+3/2)} \nonumber\\
&  \qquad {_2F_1}\bigg(1-\frac{K}{2}+l, \frac{3}{2}+L-\frac{K+N}{2}, \nonumber\\
&  \qquad \qquad L+l-K+\frac{3}{2}, \nonumber\\
&  \qquad \hspace{0.3cm} 1 - \frac{N s^\dagger \tilde C^{-1} s}{(2l-K-2)(2L-K-N-1)}\bigg)\, .
\label{glc15}
\end{eqnarray}
Using the inverse of the symmetric $2\times 2$ matrix $\tilde C$, the
relevant bilinear product evaluates to
\begin{eqnarray}
s^\dagger \tilde C^{-1} s &= \frac{\tilde C_{22}s_1^2 - 2 \tilde C_{12} s_1s_2 + \tilde C_{11}s_2^2}{\tilde C_{11} \tilde C_{22} - \tilde C_{12}^{\,2}} \nonumber\\
&= \frac{v_2^\dagger C v_2 s_1^2 - 2 v_1^\dagger C v_2 s_1s_2 + v_1^\dagger C v_1 s_2^2}{v_1^\dagger C v_1 v_2^\dagger C v_2 - (v_1^\dagger C v_2)^2} \, .
\label{glc16}
\end{eqnarray}
For our data comparison, we now consider the financial data analyzed in
II where the Algebraic--Algebraic distribution is found to describe the
data best. 
Our procedure to determine the parameters resulted
in different values for $\langle l \rangle$, $N$ and $L$ which are
listed in Tabs.~\ref{tab:FitParameterslaggr_Averaged} and
\ref{tab:FitParametersCapital_Single}. The parameter $\langle l
\rangle$ is determined by averaging over the parameter values $l$ from
all epoch distributions. We use these values in the above derived uni--
and bivariate distributions. In the case of the bivariate distribution,
we only use the values determined on the linear scale. We work out
results for the first long interval of 50 trading days with a return
horizon of 1 s, but there is no systematic difference in the results to
other intervals in the year 2014.
\begin{table}[htbp]
\begin{minipage}{1.0\linewidth}
\centering
\caption{\label{tab:FitParameterslaggr_Averaged}Averaged parameters $\langle l \rangle$, determined by logarithmic and linear fit with return horizon $\Delta t$.}
\vspace{0.3cm}
\begin{tabular}{ccc}
\toprule
fit &  $\Delta t$ & $\langle l \rangle$  \\ \hline
log &  1 s & 241.601  \\
lin &  1 s & 241.301 \\
\bottomrule
\end{tabular}
\end{minipage}%
\end{table}
\begin{table}[htbp]
\begin{minipage}{1.0\linewidth}
\centering
\caption{\label{tab:FitParametersCapital_Single}Fitting parameters for distributions of the aggregated returns on long intervals in trading days (td) and return horizon $\Delta t$ determined by logarithmic and linear fit.}\vspace{0.3cm}
\begin{tabular}{cccccc}
\toprule
interval &  fit  & $\Delta t$ & interval  & $L$ & $N$  \\
\midrule
interval \hspace{0.2cm}1 & log & 1 s & 50 td  & 338.607 & 3.123 \\
interval \hspace{0.2cm}1 & lin & 1 s & 50 td  & 339.334  & 6.051\\
\bottomrule
\end{tabular}
\end{minipage}%
\end{table}
The choice of the vectors $v_i$ is arbitrary. However we choose them
such that they are, regarding the correlations, structurally consistent
with the financial data of the market analyzed in II. As discussed in
II, the resulting multivariate distributions are dominated by the bulk
of the spectrum. The outlying large eigenvalues correspond to even
heavier tailed distributions. To test our model, we use two distinct
choices. First, we define the linear combination as a simple rotation
of three
arbitrary eigenvectors from the bulk $U_1, U_2, U_3$,
\begin{eqnarray}
v_1^\prime = \cos\psi_{12} U_1 + \sin\psi_{12} U_2\,, \nonumber\\
v_2^\prime = \cos\phi_{13} U_1 + \sin\phi_{13} U_3\,. \label{glc17}
\end{eqnarray}
Our choice of $\psi_{12} = \pi/4$ and $\phi_{13} = \pi/7$ introduces a
non--trivial dependence structure between $v_1^\prime$ and $v_2^\prime$
and therefore between $s_1$ and $s_2$.

The second choice is a very strong test for our model, as it breaks the
structural consistence. We choose an arbitrary eigenvector $U_1$ from
the bulk and keep only the first
$\lfloor K/2\rfloor$ entries, while we set
the remaining ones to zero. Then we normalize this vector to unit
length $\vert v_1^{\prime\prime} \vert = 1$.
The same is done with the second part of the vector to define
$v_2^{\prime\prime}$.
This choice differs in an important way from the simple rotation since
it tears apart the structure of the eigenvectors, which was initially
used to fit the model in II.

\begin{figure}[htbp]
\captionsetup[subfigure]{labelformat=empty}
\centering
\begin{minipage}{.5\textwidth}
\centering
\subfloat[]{\begin{overpic}[width=1.\linewidth]{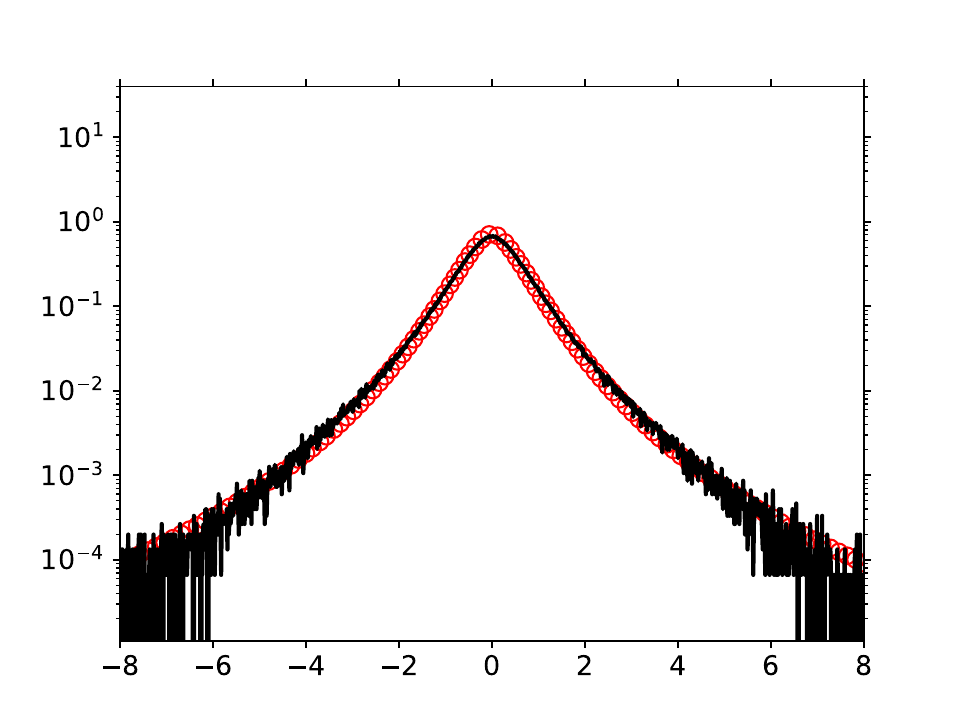}
\put(15,53){\noindent\fbox{\parbox{1.5cm}{\sffamily 50 days\\$\mathsf{\Delta t = 1\,s}$\\{\sffamily $v_1^\prime$}}}}
\put(50,0){\makebox(0,0){\small\sffamily $\mathsf{s_1}$}}
\put(2,35){\makebox(0,0){\rotatebox{90}{\small\sffamily pdf}}}
\put(78,55){\makebox(0,0){\sffamily\Large AA}}
\end{overpic}
}
\end{minipage}%
\begin{minipage}{.5\textwidth}
\centering
\subfloat[]{\begin{overpic}[width=1.\linewidth]{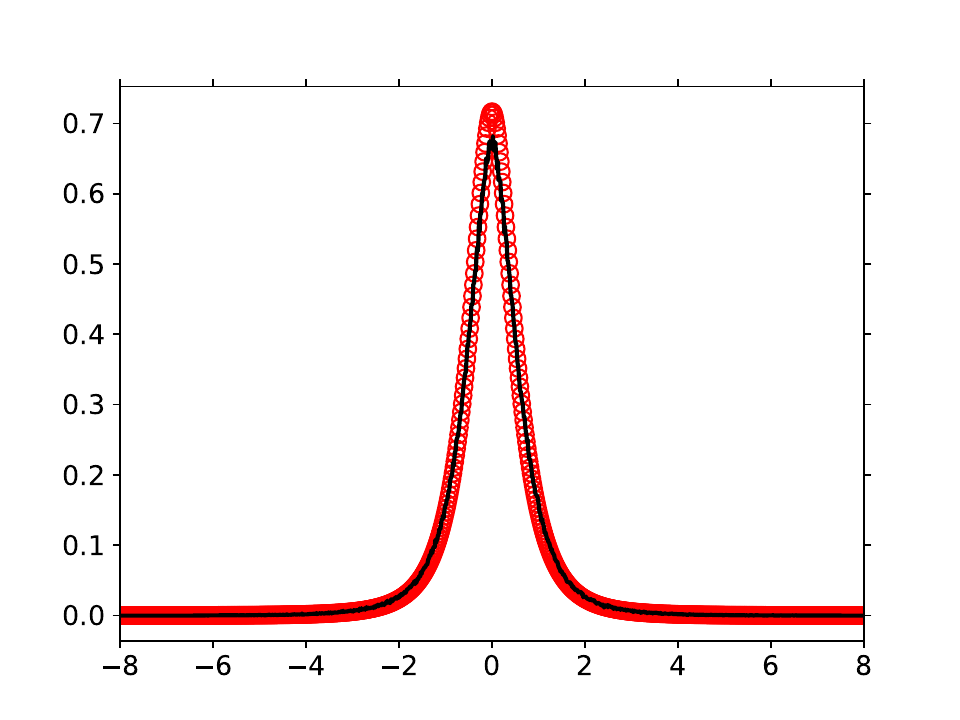}
\put(15,53){\noindent\fbox{\parbox{1.5cm}{\sffamily 50 days\\$\mathsf{\Delta t = 1\,s}$\\{\sffamily $v_1^\prime$}}}}
\put(50,0){\makebox(0,0){\small\sffamily $\mathsf{s_1}$}}
\put(2,35){\makebox(0,0){\rotatebox{90}{\small\sffamily pdf}}}
\put(78,55){\makebox(0,0){\sffamily\Large AA}}
\end{overpic}
}
\end{minipage}\\%
\begin{minipage}{.5\textwidth}
\centering
\subfloat[]{\begin{overpic}[width=1.\linewidth]{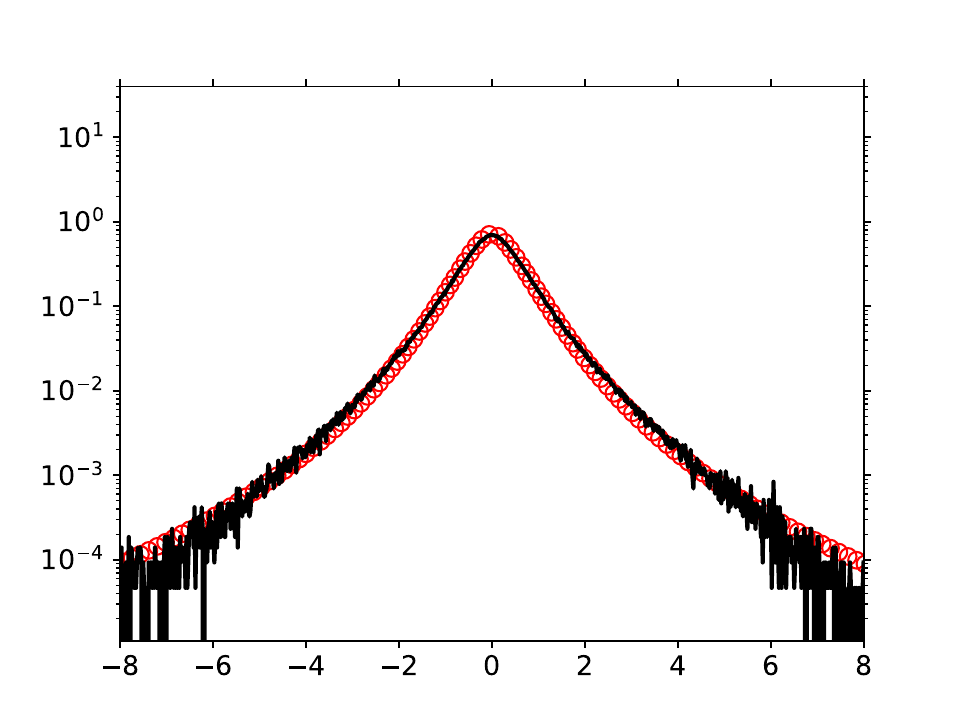}
\put(15,53){\noindent\fbox{\parbox{1.5cm}{\sffamily 50 days\\$\mathsf{\Delta t = 1\,s}$\\{\sffamily $v_2^\prime$}}}}
\put(50,0){\makebox(0,0){\small\sffamily $\mathsf{s_2}$}}
\put(2,35){\makebox(0,0){\rotatebox{90}{\small\sffamily pdf}}}
\put(78,55){\makebox(0,0){\sffamily\Large AA}}
\end{overpic}
}
\end{minipage}%
\begin{minipage}{.5\textwidth}
\centering
\subfloat[]{\begin{overpic}[width=1.\linewidth]{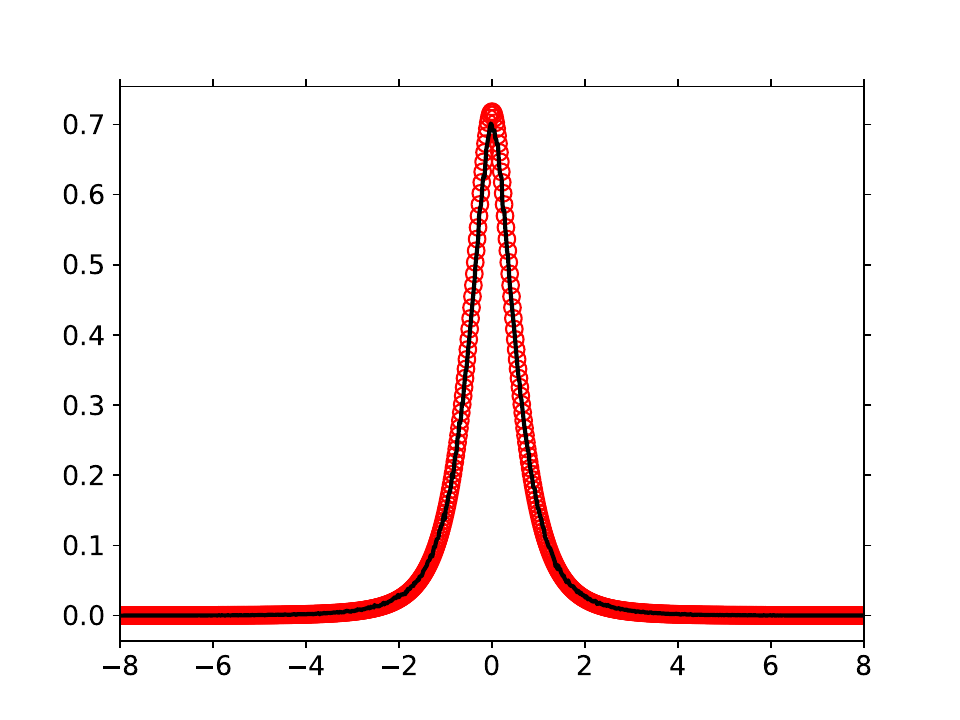}
\put(15,53){\noindent\fbox{\parbox{1.5cm}{\sffamily 50 days\\$\mathsf{\Delta t = 1\,s}$\\{\sffamily $v_2^\prime$}}}}
\put(50,0){\makebox(0,0){\small\sffamily $\mathsf{s_2}$}}
\put(2,35){\makebox(0,0){\rotatebox{90}{\small\sffamily pdf}}}
\put(78,55){\makebox(0,0){\sffamily\Large AA}}
\end{overpic}
}
\end{minipage}
\caption{\label{fig:GenLinComb_interval_Univariate_1}Empirical distributions of observables $s_1$ (top: linear combination $v_1^\prime$, $\psi_{12}=\pi/4$) and $s_2$ (bottom: linear combination $v_2^\prime$, $\phi_{13}=\pi/7$) with $\Delta t = 1\,\mathrm{s}$ (black) for the 1st long interval (50 trading days), left: on a logarithmic scale, right: on a linear scale. Model distributions $\langle p \rangle_{\mathrm{AA}}^{(1)}(s_1 \, \vert \, v_1^\dagger C v_1)$ and $\langle p \rangle_{\mathrm{AA}}^{(2)}(s_2 \, \vert \, v_2^\dagger C v_2)$ are depicted in red color.}
\end{figure}

\begin{figure}[htbp]
\captionsetup[subfigure]{labelformat=empty}
\centering
\begin{minipage}{.5\textwidth}
\centering
\subfloat[]{\begin{overpic}[width=1.\linewidth]{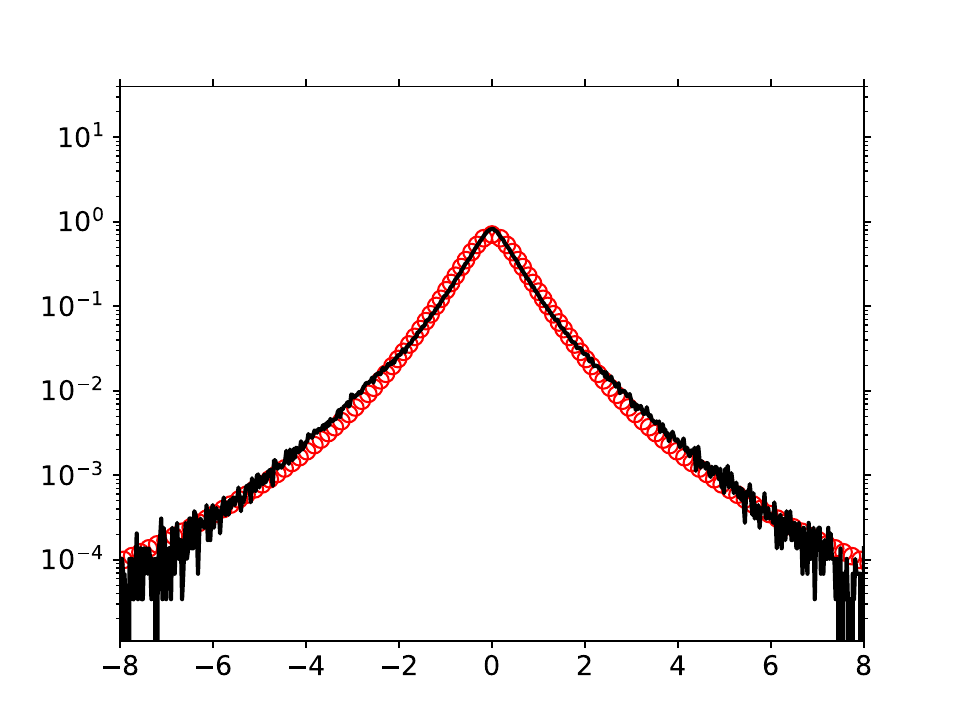}
\put(15,53){\noindent\fbox{\parbox{1.5cm}{\sffamily 50 days\\$\mathsf{\Delta t = 1\,s}$\\{\sffamily $v_1^{\prime\prime}$}}}}
\put(50,0){\makebox(0,0){\small\sffamily $\mathsf{s_1}$}}
\put(2,35){\makebox(0,0){\rotatebox{90}{\small\sffamily pdf}}}
\put(78,55){\makebox(0,0){\sffamily\Large AA}}
\end{overpic}
}
\end{minipage}%
\begin{minipage}{.5\textwidth}
\centering
\subfloat[]{\begin{overpic}[width=1.\linewidth]{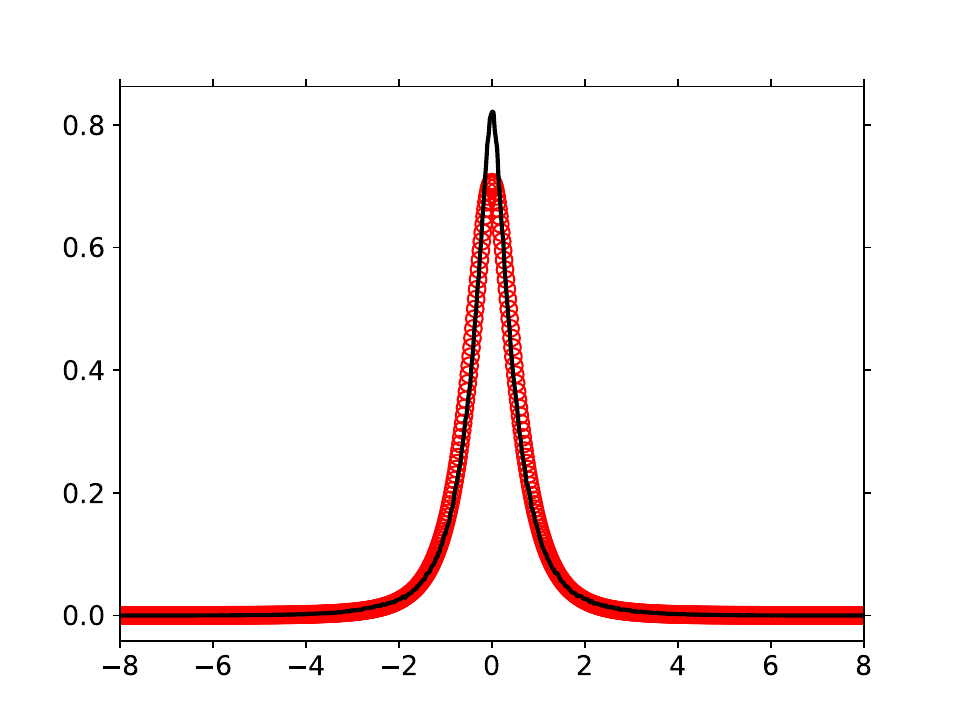}
\put(15,53){\noindent\fbox{\parbox{1.5cm}{\sffamily 50 days\\$\mathsf{\Delta t = 1\,s}$\\{\sffamily $v_1^{\prime\prime}$}}}}
\put(50,0){\makebox(0,0){\small\sffamily $\mathsf{s_1}$}}
\put(2,35){\makebox(0,0){\rotatebox{90}{\small\sffamily pdf}}}
\put(78,55){\makebox(0,0){\sffamily\Large AA}}
\end{overpic}
}
\end{minipage}\\%
\begin{minipage}{.5\textwidth}
\centering
\subfloat[]{\begin{overpic}[width=1.\linewidth]{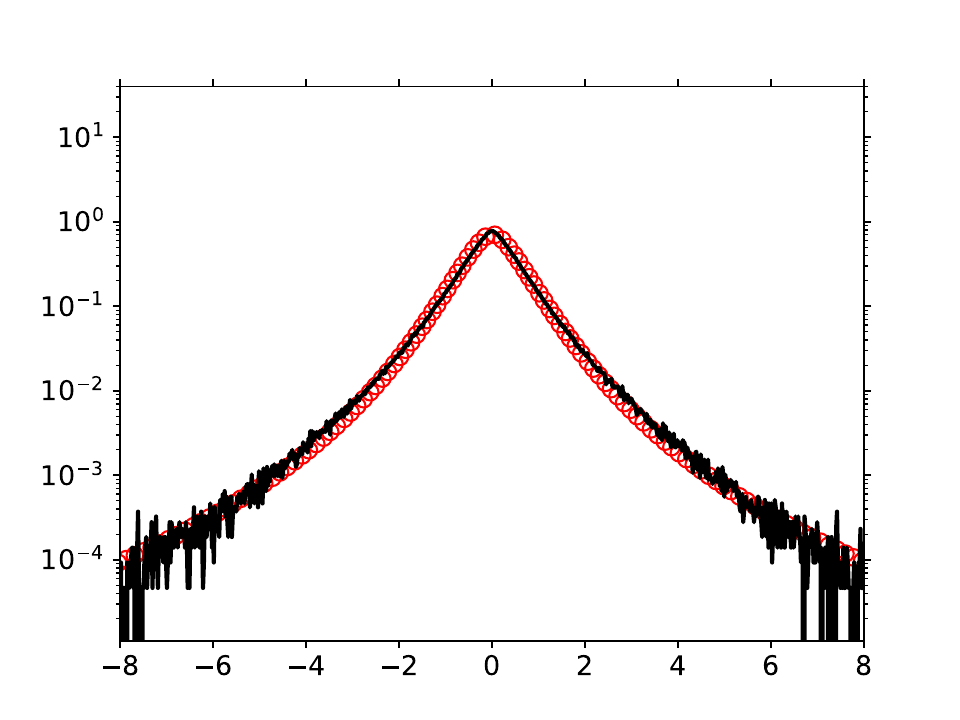}
\put(15,53){\noindent\fbox{\parbox{1.5cm}{\sffamily 50 days\\$\mathsf{\Delta t = 1\,s}$\\{\sffamily $v_2^{\prime\prime}$}}}}
\put(50,0){\makebox(0,0){\small\sffamily $\mathsf{s_2}$}}
\put(2,35){\makebox(0,0){\rotatebox{90}{\small\sffamily pdf}}}
\put(78,55){\makebox(0,0){\sffamily\Large AA}}
\end{overpic}
}
\end{minipage}%
\begin{minipage}{.5\textwidth}
\centering
\subfloat[]{\begin{overpic}[width=1.\linewidth]{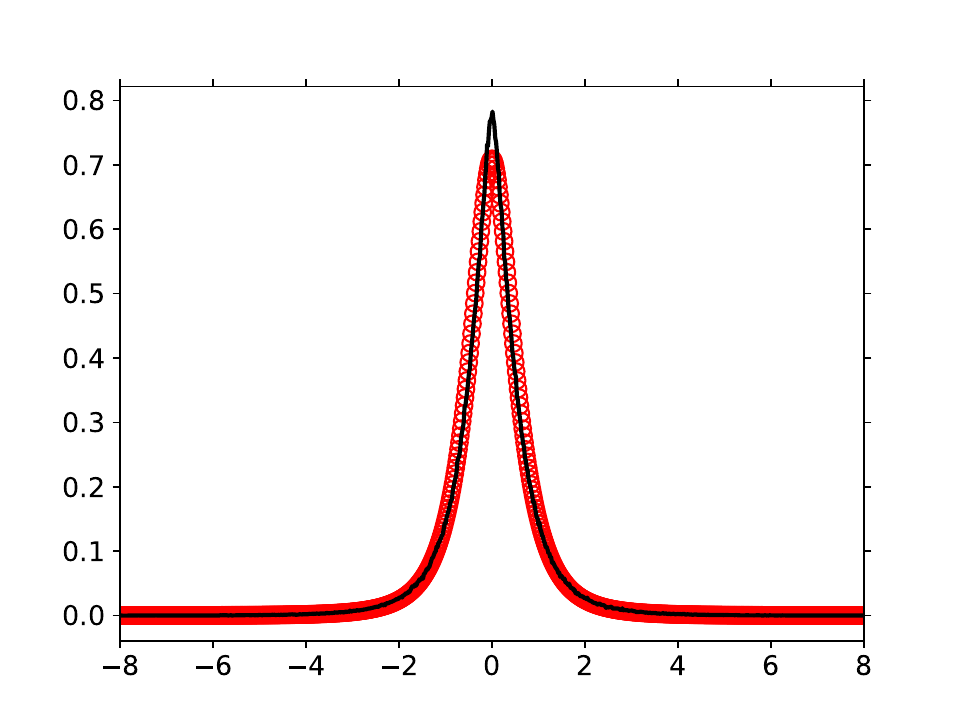}
\put(15,53){\noindent\fbox{\parbox{1.5cm}{\sffamily 50 days\\$\mathsf{\Delta t = 1\,s}$\\{\sffamily $v_2^{\prime\prime}$}}}}
\put(50,0){\makebox(0,0){\small\sffamily $\mathsf{s_2}$}}
\put(2,35){\makebox(0,0){\rotatebox{90}{\small\sffamily pdf}}}
\put(78,55){\makebox(0,0){\sffamily\Large AA}}
\end{overpic}
}
\end{minipage}
\caption{\label{fig:GenLinComb_interval_Univariate_2}Empirical distributions of observables $s_1$ (top: linear combination $v_1^{\prime\prime}$) and $s_2$ (bottom: linear combination $v_2^{\prime\prime}$) with $\Delta t = 1\,\mathrm{s}$ (black) for the 1st long interval (50 trading days), left: on a logarithmic scale, right: on a linear scale. Model distributions $\langle p \rangle_{\mathrm{AA}}^{(1)}(s_1 \, \vert \, v_1^\dagger C v_1)$ and $\langle p \rangle_{\mathrm{AA}}^{(2)}(s_2 \, \vert \, v_2^\dagger C v_2)$ are depicted in red color.}
\end{figure}

\begin{figure}[htbp]
\captionsetup[subfigure]{labelformat=empty}
\centering
\begin{minipage}{.5\textwidth}
\centering
\subfloat[]{\begin{overpic}[width=1.\linewidth]{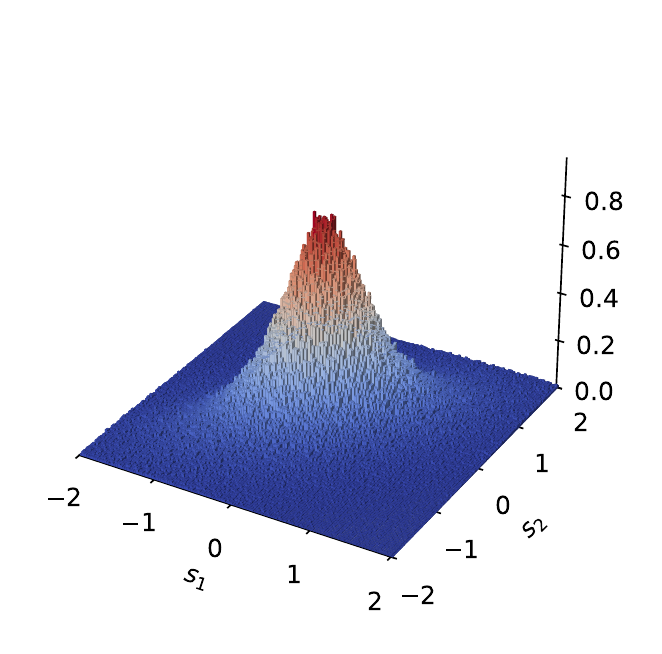}
\end{overpic}
}
\end{minipage}%
\begin{minipage}{.5\textwidth}
\centering
\subfloat[]{\begin{overpic}[width=1.\linewidth]{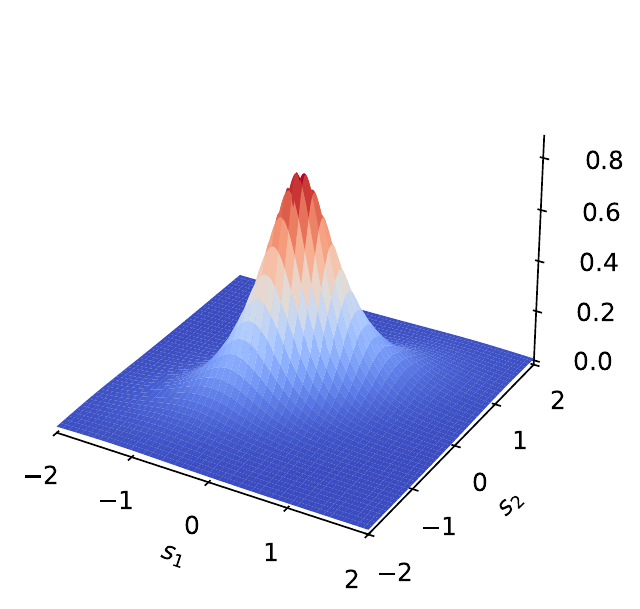}
\end{overpic}
}
\end{minipage}\\
\begin{minipage}{.5\textwidth}
\centering\vspace{-2cm}
\subfloat[]{\begin{overpic}[width=1.\linewidth]{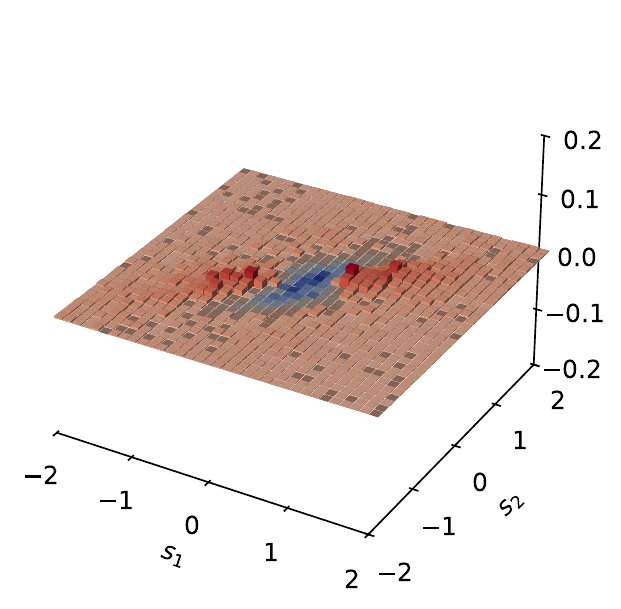}
\end{overpic}
}
\end{minipage}
\caption{\label{fig:GenLinComb_interval_Bivariate_1}Left: Bivariate empirical distributions of observables $s_1$ (linear combination $v_1^\prime$, $\psi_{12}=\pi/4$) and $s_2$ (linear combination $v_2^\prime$, $\phi_{13}=\pi/7$) with $\Delta t = 1\,\mathrm{s}$ for the 1st long interval (50 trading days). Right: Model distribution $\langle p \rangle_{\mathrm{AA}}^{(\mathrm{comb})}(s_1,s_2\,\vert\,\tilde C,\mathbb{1}_N)$. Bottom: Difference of model distribution and empirical distribution.}
\end{figure}

\begin{figure}[htbp]
\captionsetup[subfigure]{labelformat=empty}
\centering
\begin{minipage}{.5\textwidth}
\centering
\subfloat[]{\begin{overpic}[width=1.\linewidth]{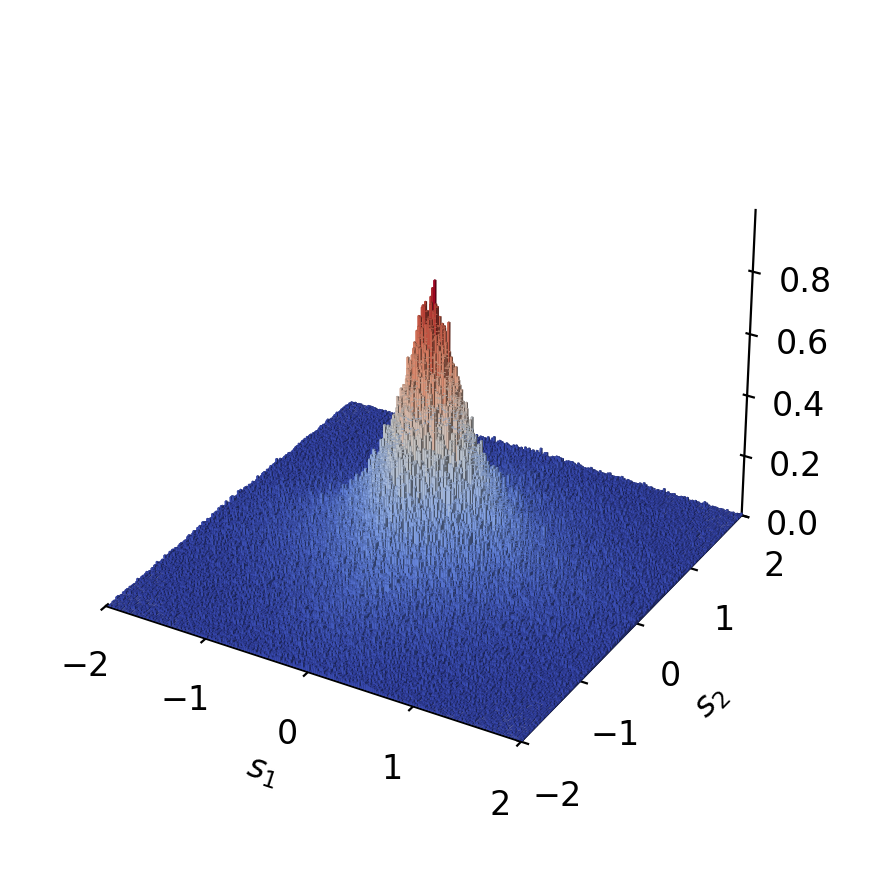}
\end{overpic}
}
\end{minipage}%
\begin{minipage}{.5\textwidth}
\centering
\subfloat[]{\begin{overpic}[width=1.\linewidth]{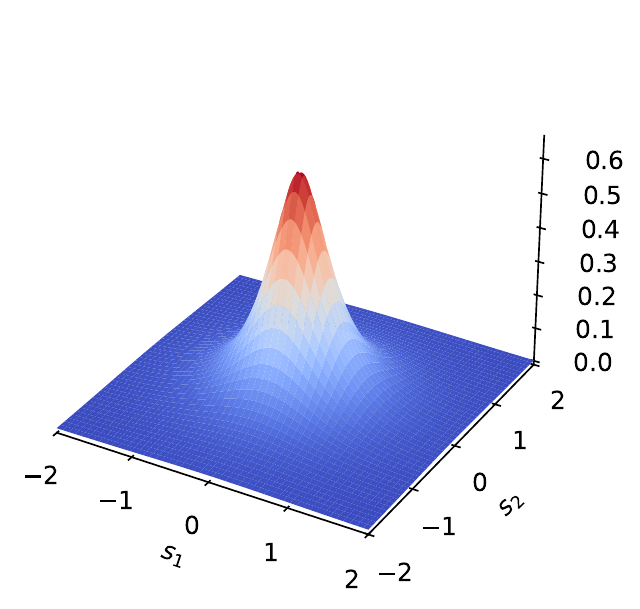}
\end{overpic}
}
\end{minipage}\\
\begin{minipage}{.5\textwidth}
\centering\vspace{-2cm}
\subfloat[]{\begin{overpic}[width=1.\linewidth]{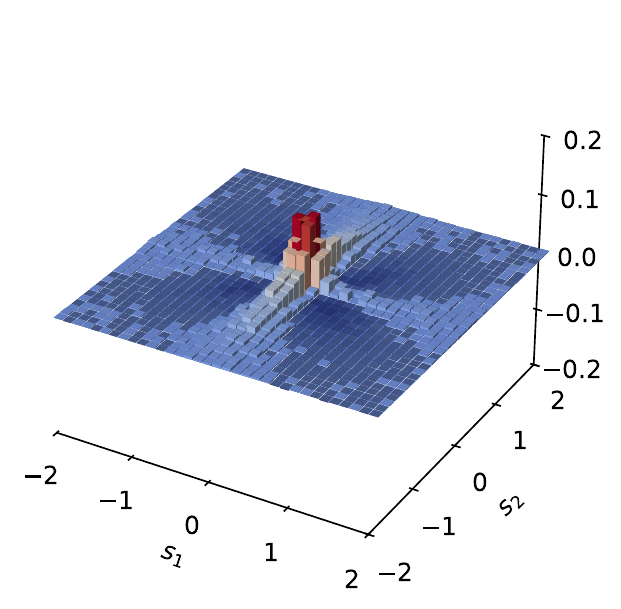}
\end{overpic}
}
\end{minipage}
\caption{\label{fig:GenLinComb_interval_Bivariate_2}Left: Bivariate empirical distributions of observables $s_1$ (linear combination $v_1^{\prime\prime}$) and $s_2$ (linear combination $v_2^{\prime\prime}$) with $\Delta t = 1\,\mathrm{s}$ for the 1st long interval (50 trading days). Right: Model distribution $\langle p \rangle_{\mathrm{AA}}^{(\mathrm{comb})}(s_1,s_2\,\vert\,\tilde C,\mathbb{1}_N)$. Bottom: Difference of model distribution and empirical distribution.}
\end{figure}

Univariate and bivariate empirical distributions are shown in  
Figs.~\ref{fig:GenLinComb_interval_Univariate_1} and
\ref{fig:GenLinComb_interval_Univariate_2}
and Figs.~\ref{fig:GenLinComb_interval_Bivariate_1} and
\ref{fig:GenLinComb_interval_Bivariate_2},
respectively. They are compared to the model distributions
(\ref{glc14}) and (\ref{glc15}).
We notice that for both types of linear combinations, $v_i^\prime$ and
$v_i^{\prime\prime}$, the univariate and bivariate model distributions
are in good agreement with the empirical ones. As expected, in
the case of the
two combinations $v_1^{\prime\prime}, v_2^{\prime\prime}$ the
differences between theoretical and empirical distributions are larger.
Importantly, we obtained these results without new parameter fits,
this demonstrates the robustness of our model. To capture really
different correlation structures, new fitting is required.


\subsection{\label{sec:MomentsModelDist}Moments of the Squared Mahalanobis Distance}

As already pointed out, the multivariate distributions in our modeling
depend on the amplitudes $r$ only via the (squared) Mahalanobis distances
\cite{MahalanobisReprint2018} $r^{\dagger} C^{-1}_\mathrm{ep} r$ and
$r^{\dagger} C^{-1} r$, respectively. As their moments are easily
empirically analyzed, it is useful for the data analysis to calculate
them in the framework of our model. Every multivariate distribution on the long
interval depending on the amplitudes $r$ has the form $f\left( r^\dagger
C^{-1} r \right)/\sqrt{\det C}$. Thus, the $\nu$--th moment of the
squared Mahalanobis distance is given by
\begin{eqnarray}
\bigl\langle ( r^{\dagger} C^{-1} r )^{\nu} \bigr\rangle =
\frac{1}{\sqrt{\det C}} \int ( r^{\dagger} C^{-1} r )^{\nu} f( r^\dagger C^{-1} r ) d[r] 
\label{mom1}
\end{eqnarray}
on the long interval and analogously on the epochs. We rewrite this as
a $w$--integral over a $\delta$--function,
\begin{eqnarray}
\bigl\langle ( r^{\dagger} C^{-1} r )^{\nu} \bigr\rangle
&= \frac{1}{\sqrt{\det C}} \int d[r] \int\limits_{0}^{\infty} dw \delta(w-r^{\dagger} C^{-1} r) w^{\nu} f(w) \nonumber\\
&= \int\limits_{0}^{\infty} dw w^{\nu} f(w) \int d[s] \delta(w-s^2) \nonumber\\
&= \frac{\pi^{K/2}}{\Gamma(K/2)}  \int\limits_{0}^{\infty} w^{K/2-1+\nu} f(w) dw \ .
\label{mom2}
\end{eqnarray}
The change of variables $r = C^{1/2}s$ is always possible because $C$
is positive definite. We further use hyperspherical coordinates and
integrate over the angles which yields the surface of the unit sphere
in $K$ dimensions. Importantly, the moments do not depend on the
correlation matrix $C$, but they depend on $D$.

The integrals (\ref{mom2}) can be done explicitly for the multivariate
amplitude distributions $p_\mathrm{Y}(r|C_\mathrm{ep})$ on the epochs,
\begin{eqnarray}
\bigl\langle ( r^{\dagger} C_{\mathrm{ep}}^{-1} r )^{\nu} \bigr\rangle_{\mathrm{ep,G}} &= \frac{2^\nu \Gamma(K/2+\nu)}{\Gamma(K/2)} \nonumber\\
\bigl\langle ( r^{\dagger} C_{\mathrm{ep}}^{-1} r )^{\nu} \bigr\rangle_{\mathrm{ep,A}} &= \frac{(2l-K-2)^\nu \Gamma(K/2+\nu)\Gamma(l-K/2-\nu)}{\Gamma(K/2)\Gamma(l-K/2)} \ ,
\label{momep}
\end{eqnarray}
where the condition $\nu<l-K/2$ holds in the algebraic case for
convergence reasons. For the multivariate
amplitude distributions $\langle p\rangle_{\mathrm{YY'}}(r|C,D)$ on the long interval, we restrict ourselves to
the Markovian case $D=\mathbb{1}_N$ and find
\begin{eqnarray}
\bigl\langle ( r^{\dagger} C^{-1} r )^{\nu} \bigr\rangle_{\mathrm{GG}} &= \left(\frac{4}{N}\right)^{\nu} \frac{\Gamma(K/2+\nu)\Gamma(N/2+\nu)}
{\Gamma(K/2)\Gamma(N/2)} \nonumber\\
\bigl\langle ( r^{\dagger} C^{-1} r )^{\nu} \bigr\rangle_{\mathrm{GA}} &= \left(\frac{2(2L-K-N-1)}{N}\right)^{\nu} \nonumber\\
&\qquad        \frac{\Gamma(K/2+\nu)\Gamma(N/2+\nu)\Gamma(L-(K+N-1)/2-\nu)}{\Gamma(K/2)\Gamma(N/2)\Gamma(L-(K+N-1)/2)} \nonumber\\
\bigl\langle ( r^{\dagger} C^{-1} r )^{\nu} \bigr\rangle_{\mathrm{AG}} &= \left(\frac{2(2l-K-2)}{N}\right)^{\nu}
\nonumber\\
&\qquad 
\frac{\Gamma(K/2+\nu)\Gamma(N/2+\nu)\Gamma(l-K/2-\nu)}{\Gamma(K/2)\Gamma(N/2)\Gamma(l-K/2)} \nonumber\\
\bigl\langle ( r^{\dagger} C^{-1} r )^{\nu} \bigr\rangle_{\mathrm{AA}} &= \left( \frac{(2l-K-2)(2L-K-N-1)}{N}\right)^{\nu} \nonumber\\
&\qquad  \frac{\Gamma(K/2+\nu)\Gamma(N/2+\nu)\Gamma(l-K/2-\nu)}{\Gamma(K/2)\Gamma(N/2)\Gamma(l-K/2)}  \nonumber \\ &\qquad\qquad \frac{\Gamma(L-(K+N-1)/2-\nu)}{\Gamma(L-(K+N-1)/2)} \ ,
\label{momli}
\end{eqnarray}
with the existence conditions $\nu<l-K/2$ and $\nu<L-(K+N-1)/2$. As
these formulae involve various $\Gamma$ functions, it is helpful to
introduce moment ratios of the form
\begin{eqnarray}
Q^{(\nu)} = \frac{\bigl\langle ( r^{\dagger} C^{-1} r )^{\nu} \bigr\rangle}{\bigl\langle r^{\dagger} C^{-1} r \bigr\rangle^\nu}
\label{mom3}
\end{eqnarray}
or similar. We consider particularly the case $\nu=2$ and arrive at
\begin{eqnarray}
Q_{\mathrm{ep,G}}^{(2)} &= \frac{K+2}{K} \nonumber\\
Q_{\mathrm{ep,A}}^{(2)} &= \frac{K+2}{K}\, \frac{2l-K-2}{2l-K-4} 
\label{momratep}
\end{eqnarray}
for the epochs and at
\begin{eqnarray}
Q_{\mathrm{GG}}^{(2)} &= \frac{(K + 2)(N + 2)}{K N} \nonumber\\
Q_{\mathrm{GA}}^{(2)} &= \frac{(K + 2)(N + 2)}{K N}\, \frac{2L-K-N-1}{2L-K-N+1} \nonumber\\
Q_{\mathrm{AG}}^{(2)} &= \frac{(K + 2)(N + 2)}{K N}\, \frac{2l-K - 2}{2l-K - 4} \nonumber\\
Q_{\mathrm{AA}}^{(2)} &= \frac{(K + 2)(N + 2)}{K N}\, \frac{2L-K-N-1}{2L-K-N+1} \, \frac{2l-K - 2}{2l-K - 4}
\label{momrattot}
\end{eqnarray}
for the long interval. These ratios have clear systematics and a
much simpler dependence on the parameters than the moments. They are
handy quantities to facilitate the parameter fixing by comparing with
their empirical values. In the Gaussian--Gaussian case, $N$ is the
only parameter and can be fixed either by fitting the distribution or
by comparing the ratios. In the other cases, combinations of both are
needed or other ratios have to be employed.

\section{\label{sec:Conclusion}Conclusions}

When analyzing data of complex systems, it is often a demanding
challenge to identify the proper observables.  Strong correlations are
typically found between the constituents or, more precisely, the
measured amplitudes.  Thus, neither the data analysis nor the
construction of models can only resort to univariate distributions.
The situation is even more involved as non--stationarity belongs to the
characterizing features of complex systems.

We carried out a new empirical analysis of non--stationarity in the
correlations by utilizing a generalized scalar product. We presented
and further extended a model for the multivariate joint
probability density functions of the measured amplitudes.  To this
end, we gave a new interpretation for Wishart--type--of approaches.  In
traditional statistics, they are used for inference, while we employed
them to model a truly existing ensemble of measured correlation
matrices in the epochs.  Choosing Gaussian and algebraic multivariate
amplitude distributions in the epochs and Gaussian and algebraic
distributions for the random model correlation matrices, we derived
four different multivariate distributions on the long interval.  Of
particular interest are the tails.  The non--stationarity fluctuations
of the correlations lift the tails when going from epochs to the long
interval.  The functional form of the distribution changes, too.  This
main result of our model is made explicit in a variety of formulae for
the data analysis which considerably extend and simplify our previous
formulae. First, we reduced the number of fit parameters in the
Algebraic formula by one, which considerably lowers the danger of ambiguous
results in the fitting routines. Second, to demonstrate that our results
have a large range of applicability, we derived multivariate distributions
for linear combinations of amplitudes from our general multivariate
distribution as obtained in II by fitting to the data. We empirically 
studied examples for uni-- and bivariate distributions and found good 
agreement without carrying out new fits. This strongly corroborates the
robustness of our model. Third, we calculated moments and ratios of 
moments of the Mahalanobis distance which will also facilitate the data
analysis.

In the forthcoming study II, we apply our findings to a correlated
financial market, further applications to other complex systems are
planned.

\section*{Acknowledgment}

We thank Shanshan Wang for fruitful discussions.

\clearpage
\markboth{References}{References}

\newcommand{\newblock}{}

\printbibliography[heading=none]%
%

\newpage

\begin{appendices}
\markboth{Appendix}{Appendix}
\section{\label{app:FitParameterReduction}Importance of Parameter Reduction in Least--Squares Fitting}
Upon the referee's request we provide some basic explanations
of standard issues in multi--parameter fitting. We refer to
the common literature, such
as~\cite{BevingtonRobinson2003,Taylor1997,Lyons1991,HansenPereyraScherer2013,PressEtAl2007,NarskyPorter2014}. To
avoid confusion for the reader, we emphasize that the
discussion in the sequel refers exclusively to the
multi--parameter fitting of empirical data in II. In
Sec.~\ref{sec:GeneralLinearComb} of the present paper I, no new
fits are carried out, only the results of the fits in II are
used.

Consider a function of a variable, say $r$, which depends on
$f$ parameters $\alpha_1,\ldots,\alpha_f$ which has to be fitted
to experimental or empirical data. The least--squares fitting
procedures amount to finding a minimum of the sum $\chi^2$ of
the squared residuals in the $f$ dimensional parameter
space. The higher the number of parameters, the more likely
is the occurrence of several (local) minima in the parameter
space. The challenge is to find a global
minimum~\cite{BevingtonRobinson2003,Taylor1997,Lyons1991,HansenPereyraScherer2013,PressEtAl2007,NarskyPorter2014}. Unfortunately,
the multiple local minima are often close together in
parameter space. This proximity increases the risk of
convergence to a false minimum depending heavily on the
initial conditions of the fitting algorithm. In
high--dimensional parameter spaces, there is a significantly
greater risk of becoming trapped in local minima, which often
leads to ambiguous results.  Thus, it is highly desirable to reduce the number of
parameters wherever possible --- preferably in a physically or
statistically well motivated manner.

To illustrate this we revisit here the fits carried out in
II, but importantly in contrast to II \textit{without}
reducing the number of fit parameters. We recall that formulae
(\ref{eq:GauVerR1beta}) and (\ref{eq:GauVerR1Beta}) were used
to reduce the number of fit parameters on the epochs from two
to one and on the long interval from three to two,
respectively.  Hence, we employ here for the fits on the
epochs the distribution~(\ref{eq:RotRescAlgebraic}), but undo
the parameter reduction: the distribution for the present
purpose depends on $l_{\mathrm{rot}}$ and $m$, where $l$ and
$l_{\mathrm{rot}}$ have the linear relation (\ref{eq:laggr}).
On the large interval, we employ Eq.~(114) in
\cite{Guhr_2021}, which, after $l_{\mathrm{rot}}$ and $m$ have
been fixed by epoch fits, depends on the three parameters $N$,
$L$ and $M$, instead of only two in II. For convenience, we use
$L_{\mathrm{rot}}$ where $L$ and $L_{\mathrm{rot}}$ have the
linear relation (\ref{eq:Laggr}). To address the distributions
of the aggregated returns on the long intervals $\langle p
\rangle_{\mathrm{AA}}^{(\mathrm{aggr})} (\widetilde{r})$ and
the epochs $p_{\mathrm{A}}^{(\mathrm{aggr})} (\widetilde{r})$,
respectively, we set the
eigenvalues $\Lambda_k$ and $\Lambda_{\mathrm{ep},k}$ to one.

We proceed in two steps. First, we fit
the empirical distributions of the aggregated returns on each of the epochs and find the parameters $l_{\mathrm{rot}}$ and $m$.  For
the long intervals, we use the mean values $\langle
l_{\mathrm{rot}} \rangle$ and $\langle m \rangle$, which are
obtained by averaging over the fit parameters of all 250
epochs on a logarithmic and linear scale, see
Tab.~\ref{tab:FitParameters_lrot_m_Averaged}.  Of course, in
II we only had to determine and use $\langle l_{\mathrm{rot}}
\rangle$.
\begin{table}[htbp]
\begin{minipage}{1.0\linewidth}
\centering
\caption{\label{tab:FitParameters_lrot_m_Averaged}Averaged parameters $\langle l_\mathrm{rot} \rangle$ and $\langle m \rangle$ determined by logarithmic and linear fit with return horizon $\Delta t = 1\,$s.}
\vspace{0.3cm}
\begin{tabular}{ccccc}
\toprule
fit &  $\Delta t$ & $\langle l_\mathrm{rot} \rangle$  & $\langle m \rangle$   \\ \hline
log &  1\,s & 2.77  & 2.76    \\
lin &  1\,s & 1.98  & 1.28  \\
\bottomrule
\end{tabular}
\end{minipage}%
\end{table}%
Second, we fit the distributions of the aggregated
returns on some of the long intervals to the empirical
data.
\begin{figure}[htbp]
\captionsetup[subfigure]{labelformat=empty}
\centering
\begin{minipage}{.5\textwidth}
\centering
\subfloat[]{\begin{overpic}[width=1.\linewidth]{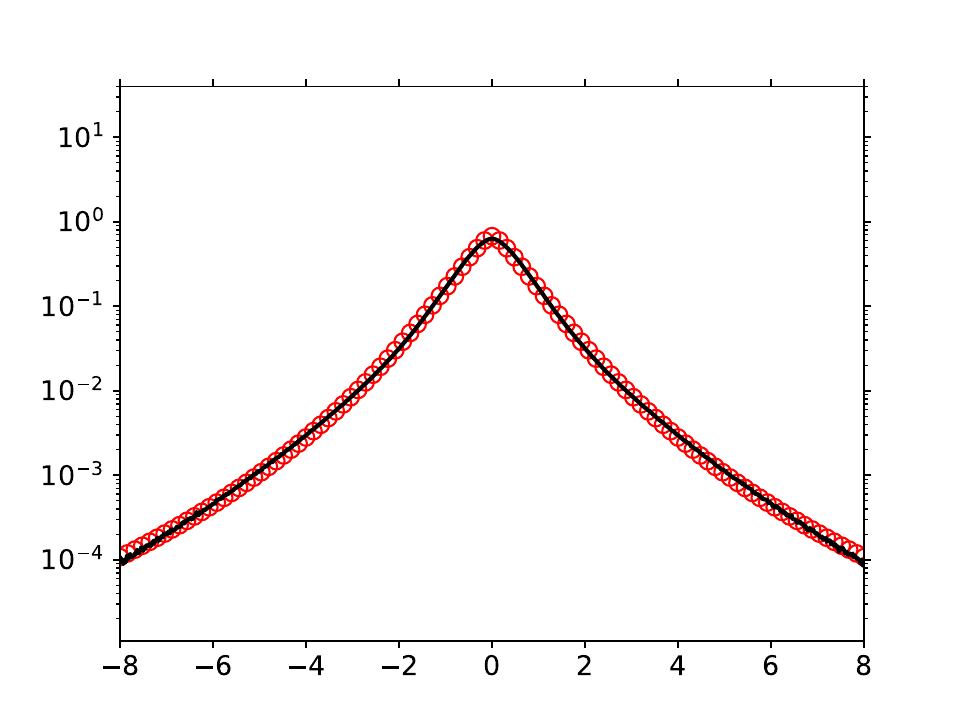}
\put(15,56){\noindent\fbox{\parbox{2.5cm}{\sffamily long interval 1}}}
\put(38,20){\noindent\fbox{\parbox{1.8cm}{\sffamily 25 days\\$\mathsf{\Delta t = 1\,s}$}}}
\put(50,0){\makebox(0,0){\small\sffamily aggregated return}}
\put(2,35){\makebox(0,0){\rotatebox{90}{\small\sffamily pdf}}}
\put(78,57){\makebox(0,0){\sffamily\Large AA}}
\end{overpic}
}
\end{minipage}%
\begin{minipage}{.5\textwidth}
\centering
\subfloat[]{\begin{overpic}[width=1.\linewidth]{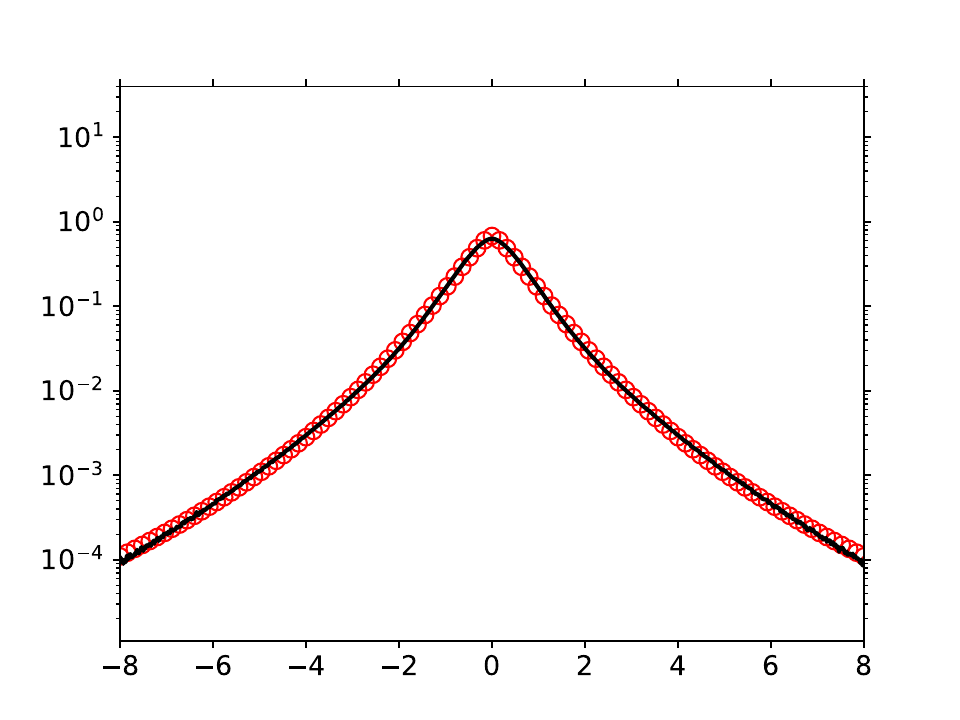}
\put(15,56){\noindent\fbox{\parbox{2.5cm}{\sffamily long interval 1}}}
\put(38,20){\noindent\fbox{\parbox{1.8cm}{\sffamily 25 days\\$\mathsf{\Delta t = 1\,s}$}}}
\put(50,0){\makebox(0,0){\small\sffamily aggregated return}}
\put(2,35){\makebox(0,0){\rotatebox{90}{\small\sffamily pdf}}}
\put(78,57){\makebox(0,0){\sffamily\Large AA}}
\end{overpic}
}
\end{minipage}\\%
\begin{minipage}{.5\textwidth}
\centering
\subfloat[]{\begin{overpic}[width=1.\linewidth]{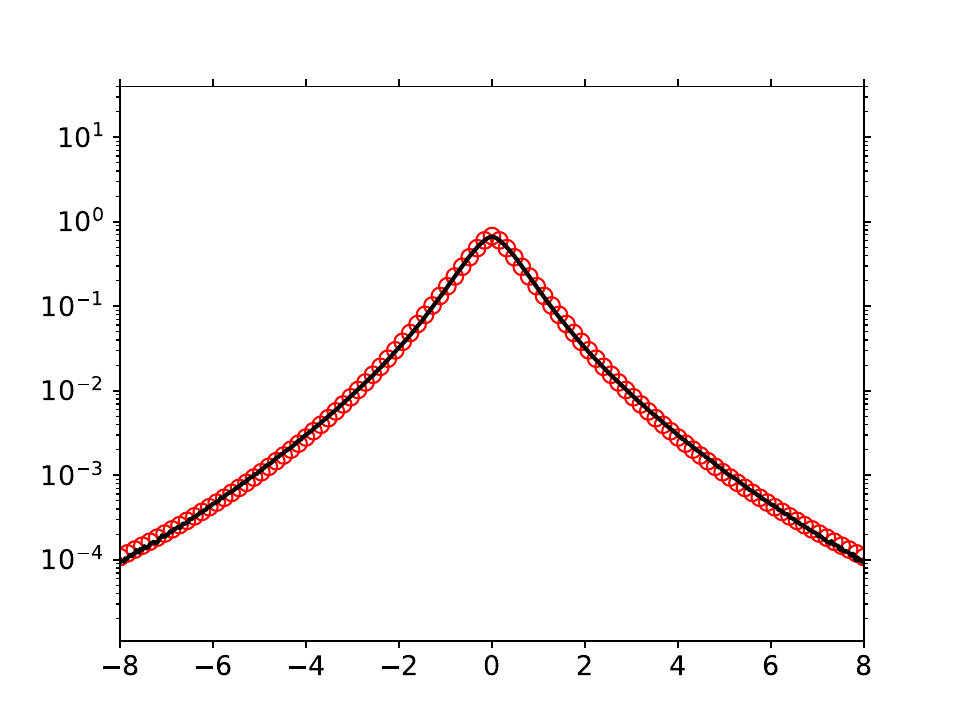}
\put(15,56){\noindent\fbox{\parbox{2.7cm}{\sffamily long interval 10}}}
\put(38,20){\noindent\fbox{\parbox{1.8cm}{\sffamily 25 days\\$\mathsf{\Delta t = 1\,s}$}}}
\put(50,0){\makebox(0,0){\small\sffamily aggregated return}}
\put(2,35){\makebox(0,0){\rotatebox{90}{\small\sffamily pdf}}}
\put(78,57){\makebox(0,0){\sffamily\Large AA}}
\end{overpic}
}
\end{minipage}%
\begin{minipage}{.5\textwidth}
\centering
\subfloat[]{\begin{overpic}[width=1.\linewidth]{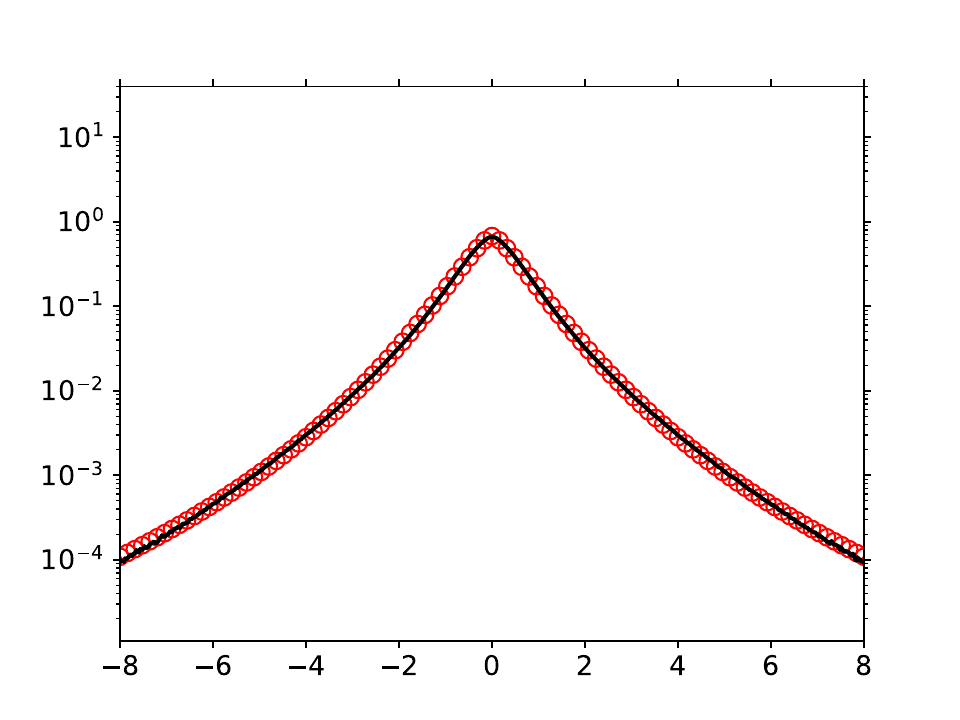}
\put(15,56){\noindent\fbox{\parbox{2.7cm}{\sffamily long interval 10}}}
\put(38,20){\noindent\fbox{\parbox{1.8cm}{\sffamily 25 days\\$\mathsf{\Delta t = 1\,s}$}}}
\put(50,0){\makebox(0,0){\small\sffamily aggregated return}}
\put(2,35){\makebox(0,0){\rotatebox{90}{\small\sffamily pdf}}}
\put(78,57){\makebox(0,0){\sffamily\Large AA}}
\end{overpic}
}
\end{minipage} 
\caption{Empirical (black) and model (red, Algebraic–Algebraic) distributions of aggregated returns with $\Delta t = 1\,\mathrm{s}$ on a logarithmic scale for long intervals (25 trading days). Top: long interval 1; bottom: long interval 10. Fit parameters are given in Tab.~\ref{tab:FitParameters_Lrot_M_Chi2_Averaged}.}
\label{fig:FitsDifferentLrotM_LogScale}
\end{figure}%
\begin{figure}[htbp]
\captionsetup[subfigure]{labelformat=empty}
\centering
\begin{minipage}{.5\textwidth}
\centering
\subfloat[]{\begin{overpic}[width=1.\linewidth]{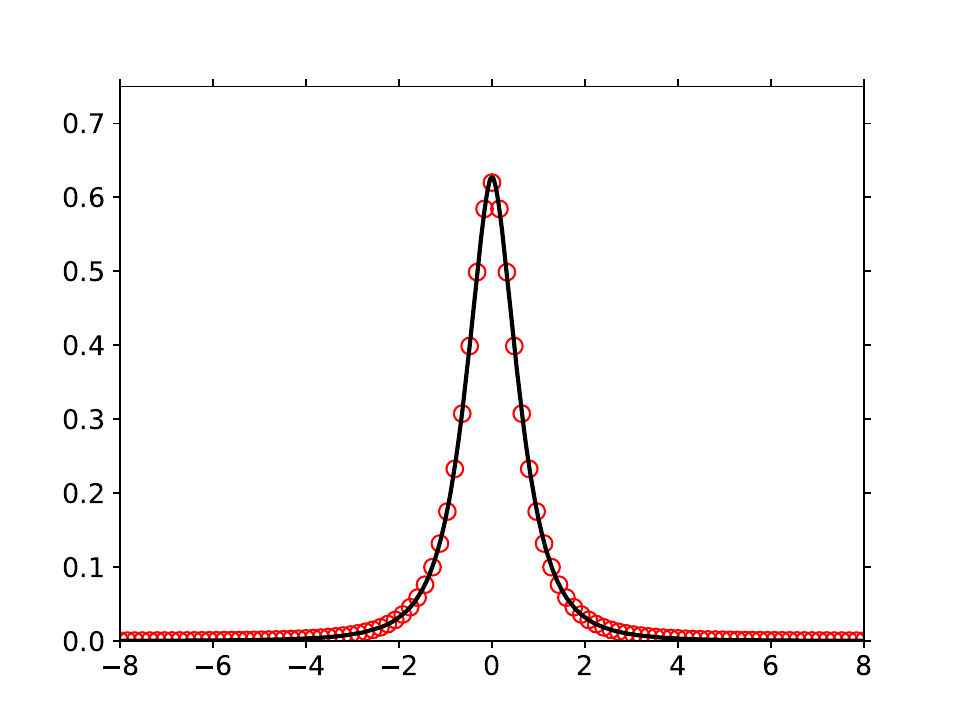}
\put(14,58){\noindent\fbox{\parbox{2.5cm}{\sffamily long interval 2}}}
\put(16,20){\noindent\fbox{\parbox{1.8cm}{\sffamily 25 days\\$\mathsf{\Delta t = 1\,s}$}}}
\put(50,0){\makebox(0,0){\small\sffamily aggregated return}}
\put(2,35){\makebox(0,0){\rotatebox{90}{\small\sffamily pdf}}}
\put(78,57){\makebox(0,0){\sffamily\Large AA}}
\end{overpic}
}
\end{minipage}%
\begin{minipage}{.5\textwidth}
\centering
\subfloat[]{\begin{overpic}[width=1.\linewidth]{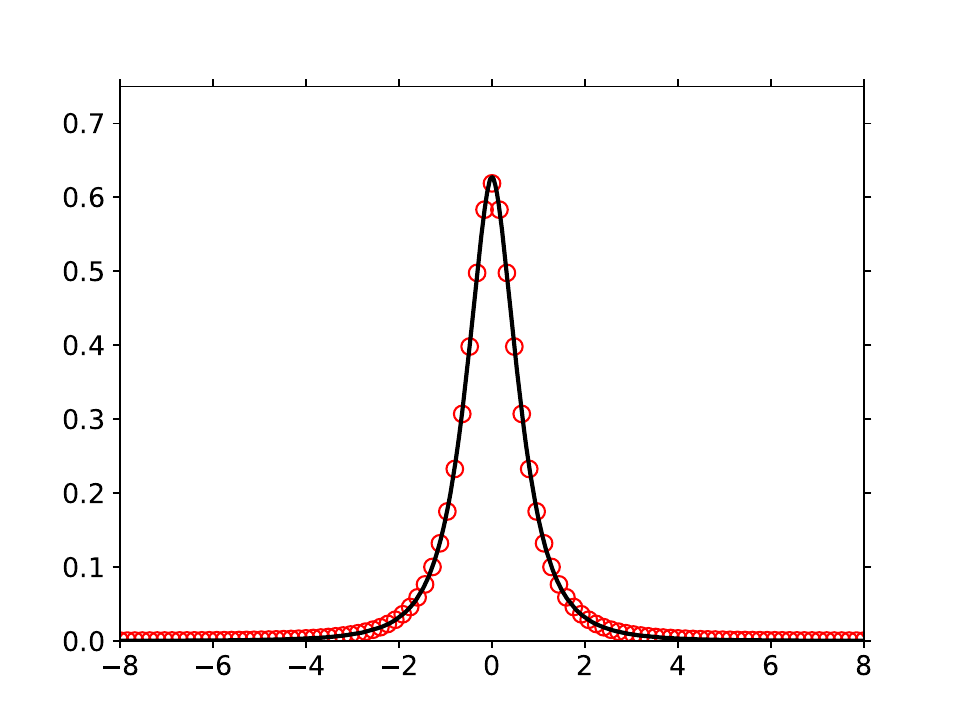}
\put(14,58){\noindent\fbox{\parbox{2.5cm}{\sffamily long interval 2}}}
\put(16,20){\noindent\fbox{\parbox{1.8cm}{\sffamily 25 days\\$\mathsf{\Delta t = 1\,s}$}}}
\put(50,0){\makebox(0,0){\small\sffamily aggregated return}}
\put(2,35){\makebox(0,0){\rotatebox{90}{\small\sffamily pdf}}}
\put(78,57){\makebox(0,0){\sffamily\Large AA}}
\end{overpic}
}
\end{minipage}\\%
\begin{minipage}{.5\textwidth}
\centering
\subfloat[]{\begin{overpic}[width=1.\linewidth]{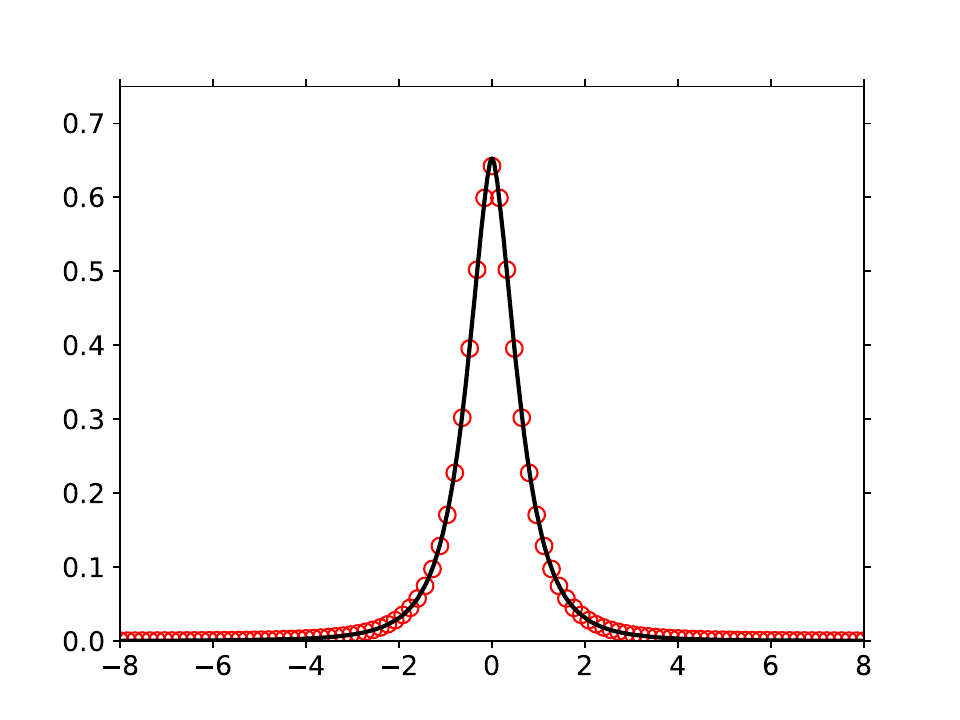}
\put(14,58){\noindent\fbox{\parbox{2.5cm}{\sffamily long interval 6}}}
\put(16,20){\noindent\fbox{\parbox{1.8cm}{\sffamily 25 days\\$\mathsf{\Delta t = 1\,s}$}}}
\put(50,0){\makebox(0,0){\small\sffamily aggregated return}}
\put(2,35){\makebox(0,0){\rotatebox{90}{\small\sffamily pdf}}}
\put(78,57){\makebox(0,0){\sffamily\Large AA}}
\end{overpic}
}
\end{minipage}%
\begin{minipage}{.5\textwidth}
\centering
\subfloat[]{\begin{overpic}[width=1.\linewidth]{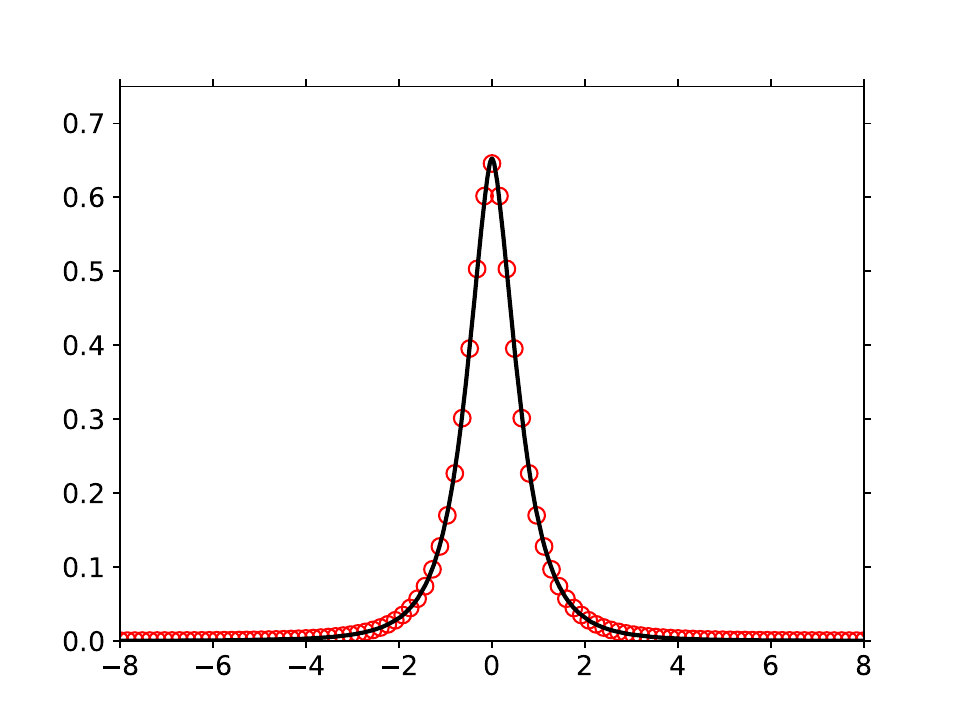}
\put(14,58){\noindent\fbox{\parbox{2.5cm}{\sffamily long interval 6}}}
\put(16,20){\noindent\fbox{\parbox{1.8cm}{\sffamily 25 days\\$\mathsf{\Delta t = 1\,s}$}}}
\put(50,0){\makebox(0,0){\small\sffamily aggregated return}}
\put(2,35){\makebox(0,0){\rotatebox{90}{\small\sffamily pdf}}}
\put(78,57){\makebox(0,0){\sffamily\Large AA}}
\end{overpic}
}
\end{minipage}
\caption{Empirical (black) and model (red, Algebraic–Algebraic) distributions of aggregated returns with $\Delta t = 1\,\mathrm{s}$ on a linear scale for long intervals (25 trading days). Top: long interval 2; bottom: long interval 6. Fit parameters are given in Tab.~\ref{tab:FitParameters_Lrot_M_Chi2_Averaged}.}
\label{fig:FitsDifferentLrotM_LinScale}
\end{figure}
\begin{table}[htbp]
\begin{minipage}{1.0\linewidth}
\centering
\caption{\label{tab:FitParameters_Lrot_M_Chi2_Averaged}Parameters $N$, $L_\mathrm{rot}$ and $M$ corresponding to Figs.~\ref{fig:FitsDifferentLrotM_LogScale} and \ref{fig:FitsDifferentLrotM_LinScale}.}
\vspace{0.3cm}
\begin{tabular}{cccccccc}
\toprule
fit & $\Delta t$ & long interval & $N$ & $L_\mathrm{rot}$ & $M$ & $\chi^2$ \\
\hline
log & 1\,s & long interval 1 & 3.20 & 168.85 & 312.68 & 0.004 \\
log & 1\,s & long interval 1 & 3.20 & 43.79 & 77.60 & 0.004 \\
log & 1\,s & long interval 10 & 3.16 & 168.85 & 312.26 & 0.003 \\
log & 1\,s & long interval 10 & 3.16 & 95.17 & 174.47 & 0.003 \\
lin & 1\,s & long interval 2 & 9.31 & 80.75 & 146.66 & $1.12~\cdot10^{-6}$ \\
lin & 1\,s & long interval 2 & 9.31 & 50.32 & 88.13 & $1.23~\cdot10^{-6}$ \\
lin & 1\,s & long interval 6 & 6.78 & 159.84 & 302.36 & $1.67~\cdot10^{-6}$ \\
lin & 1\,s & long interval 6 & 6.78 & 65.79 & 119.01 & $2.06~\cdot10^{-6}$ \\
\bottomrule
\end{tabular}
\end{minipage}%
\end{table}%
To emphasize the ambiguity, we fit $N$ only once and
then fix it, and fit only $L_{\mathrm{rot}}$ and
$M$. As seen, the distributions on
different long intervals are
hardly distinguishable in
Figs.~\ref{fig:FitsDifferentLrotM_LogScale}~and~\ref{fig:FitsDifferentLrotM_LinScale},
while the corresponding parameters $L_{\mathrm{rot}}$
and $M$ in
Tab.~\ref{tab:FitParameters_Lrot_M_Chi2_Averaged} are
very different. Even large variations of the values
for $L_{\mathrm{rot}}$ and $M$, yield the same quality
of fit, as measured by $\chi^2$.

\begin{figure}[htbp]
\captionsetup[subfigure]{labelformat=empty}
\centering
\begin{minipage}{.5\textwidth}
\centering
\subfloat[]{\begin{overpic}[width=1.\linewidth]{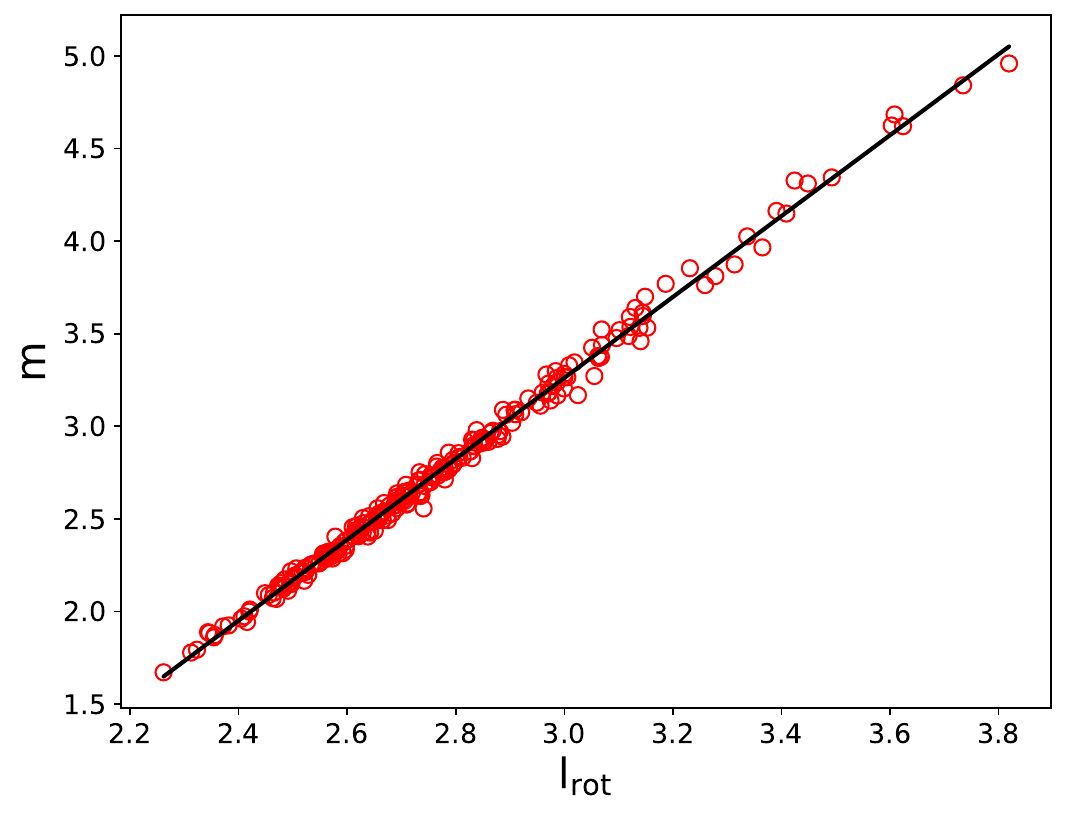}
\end{overpic}
}
\end{minipage}%
\begin{minipage}{.5\textwidth}
\centering
\subfloat[]{\begin{overpic}[width=1.\linewidth]{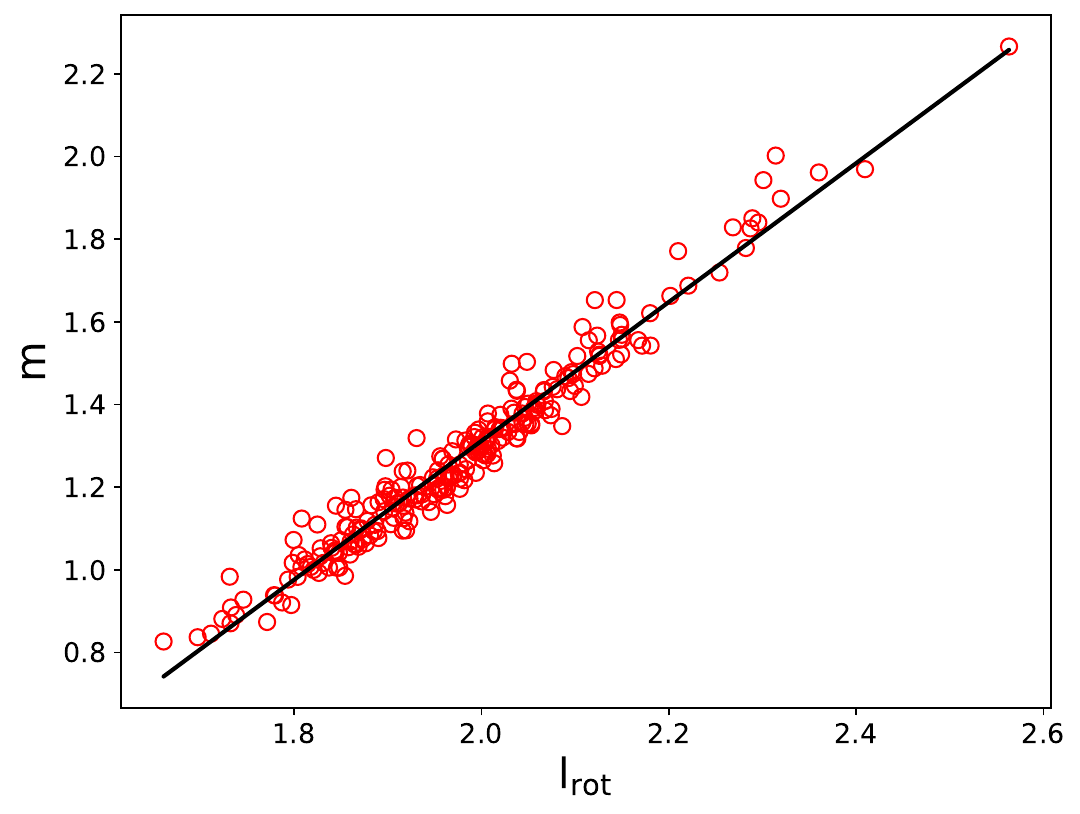}
\end{overpic}
}
\end{minipage}
\caption{Scatter plots of fit parameters $l_{\mathrm{rot}}$ and $m$ determined by fits on logarithmic (left) and linear (right) scales, computed over the epochs with a return horizon $\Delta t = 1\,\mathrm{s}$.}
\label{fig:ScatterPlotslrotm}
\end{figure}

\begin{figure}[htbp]
\captionsetup[subfigure]{labelformat=empty}
\centering
\begin{minipage}{.5\textwidth}
\centering
\subfloat[]{\begin{overpic}[width=1.\linewidth]{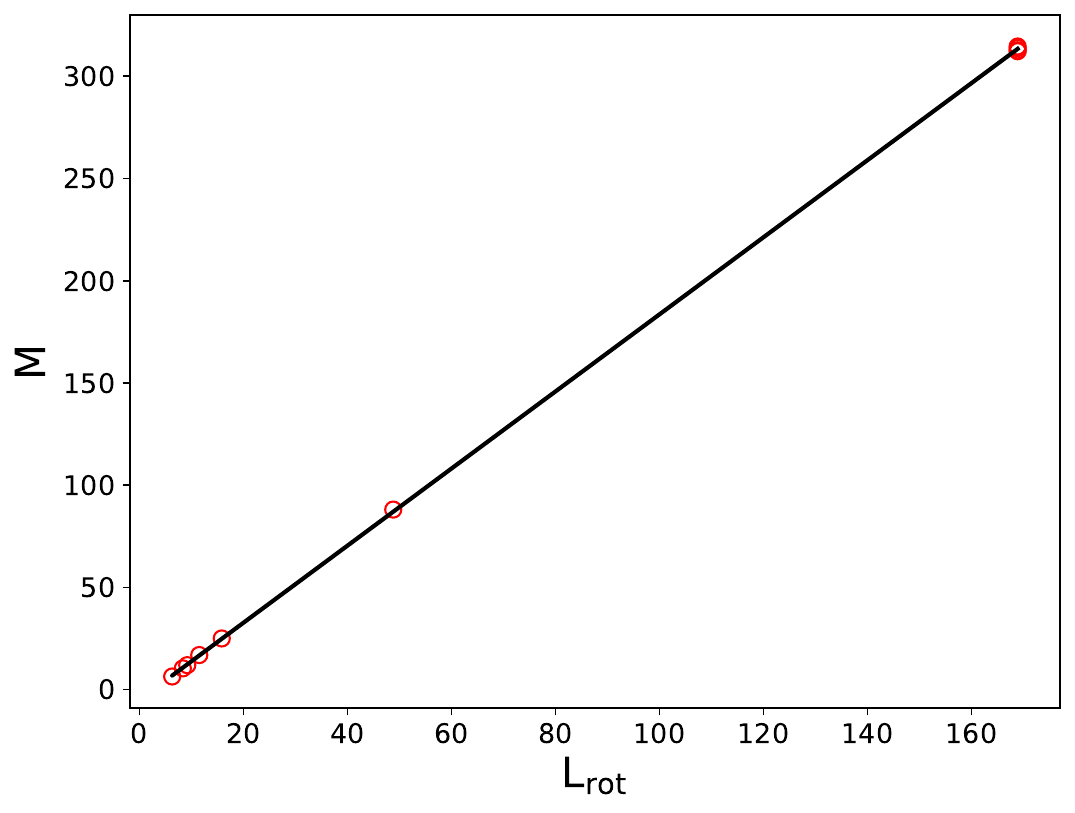}
\end{overpic}
}
\end{minipage}%
\begin{minipage}{.5\textwidth}
\centering
\subfloat[]{\begin{overpic}[width=1.\linewidth]{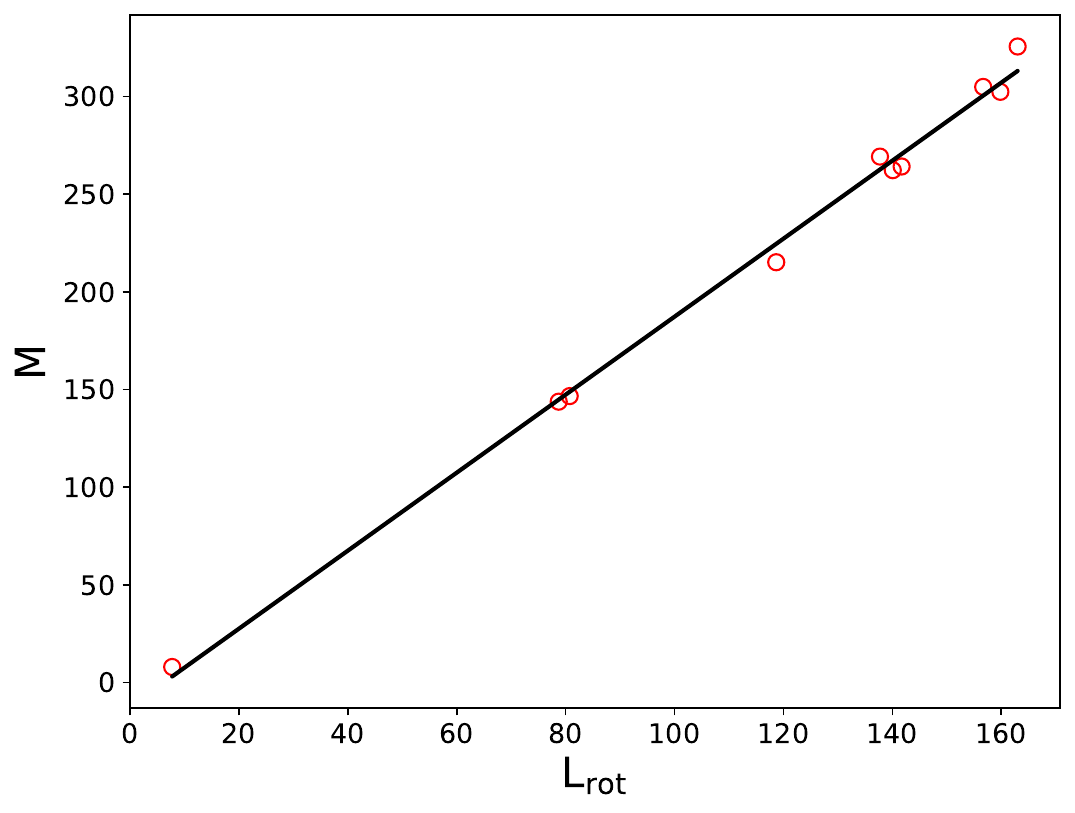}
\end{overpic}
}
\end{minipage}
\caption{Scatter plots of fit parameters $L_{\mathrm{rot}}$ and $M$ determined by fits on logarithmic (left) and linear (right) scales, computed over long intervals of 25 trading days with a return horizon $\Delta t = 1\,\mathrm{s}$.}
\label{fig:ScatterPlotsLrotM}
\end{figure}

The reason for this ambiguity is precisely that equations (\ref{eq:GauVerR1beta}) and (\ref{eq:GauVerR1Beta}) connect $l$ or equivalently $l_{\mathrm{rot}}$ and $m$, and $L$ or equivalently $L_{\mathrm{rot}}$ and $M$, respectively. As
Figs.~\ref{fig:ScatterPlotslrotm} and ~\ref{fig:ScatterPlotsLrotM} show, the obtained
value pairs $(l_{\mathrm{rot}}, m)$ for all epochs and $(L_{\mathrm{rot}}, M)$
for all long
intervals lie on a straight line, determined by
equations (\ref{eq:GauVerR1beta}) and (\ref{eq:GauVerR1Beta}), respectively.

In summary, we demonstrated in this
Appendix the importance of parameter reduction for fitting
procedures. Here, we deliberately relinquished the
formulae (\ref{eq:GauVerR1beta}) and
(\ref{eq:GauVerR1Beta}) which we used in II for parameter
reduction. Thus, the fits here led to ambiguous results.
Furthermore, we demonstrated that plotting the obtained
values for fit parameters versus each other can help to
identify mutual relations, in our case this only confirmed
formulae (\ref{eq:GauVerR1beta}) and
(\ref{eq:GauVerR1Beta}).

\end{appendices}

\end{document}